\documentstyle[11pt,aaspp4,psfig]{article}
\def\be{\begin{equation}}
\def\ee{\end{equation}}

\def\lsim{\lower 2pt \hbox{$\, \buildrel {\scriptstyle <}\over
         {\scriptstyle \sim}\,$}}
\newcommand\gsim{\buildrel > \over \sim}
\begin{document}
\newcommand{\figureout}[3]{\psfig{figure=#1,width=7in,angle=#2} 
   \figcaption{#3} }

\title{Pulsar Polar Cap Heating and Surface Thermal X-Ray\\ 
 Emission II. Inverse Compton Radiation Pair Fronts}

\author{Alice K. Harding\altaffilmark{1} and Alexander G. Muslimov\altaffilmark{2}
}
\altaffiltext{1}{Laboratory of High Energy Astrophysics, 
NASA/Goddard Space Flight Center, Greenbelt, MD 20771}

\altaffiltext{2}{Emergent Information Technologies, 
Inc., Space Sciences Sector, Upper Marlboro, MD 20774}  

\begin{abstract}

We investigate the production of electron-positron pairs by inverse Compton scattered
(ICS) photons above a pulsar polar cap (PC) and calculate surface heating by returning 
positrons. This paper is a continuation of our self-consistent treatment of acceleration, 
pair dynamics and electric field screening above pulsar PCs.  We calculate the altitude of 
the inverse Compton pair formation fronts, the flux of returning positrons and present 
the heating efficiencies and X-ray luminosities.  We revise pulsar death lines implying 
cessation of pair formation, and present them in surface magnetic field-period space.  
We find that virtually all known radio pulsars are capable of producing pairs by resonant 
and non-resonant ICS photons radiated by particles accelerated above the PC in a pure 
star-centered dipole field, so that our ICS pair death line coincides with empirical 
radio pulsar death.  Our calculations show that ICS pairs are able to screen the 
accelerating electric field only for high PC surface temperatures and magnetic 
fields.  We argue that such screening at ICS pair fronts occurs locally, slowing but not 
turning off acceleration of particles until screening can occur at a curvature 
radiation (CR) pair front at higher altitude.  In the case where no screening 
occurs above the PC surface, we anticipate that the pulsar $\gamma $-ray luminosity 
will be a substantial fraction of its spin-down luminosity. The X-ray luminosity resulting 
from PC heating by ICS pair fronts is significantly lower than the PC heating luminosity 
from CR pair fronts, which dominates for most pulsars.  PC heating from 
ICS pair fronts is highest in millisecond pulsars, which cannot produce CR pairs, and 
may account for observed thermal X-ray components in the spectra of these old pulsars.

\end{abstract} 


\pagebreak
  
\section{INTRODUCTION}

In the last several years, the basic model of particle acceleration above a 
pulsar polar cap (PC) has been undergoing significant revision.  Sturrock (1971),
Ruderman \& Sutherland (1975) and Arons \& Scharlemann (1979) originally proposed
that particles are accelerated by an induced electric field, producing curvature 
radiation (CR) photons which create electron-positron pairs in the strong magnetic 
field.  The pairs short out the electric
field above a pair-formation front (PFF), self-limiting the acceleration.  
In the process of screening or shorting-out the electric field, some fraction of 
the positrons is accelerated back toward the PCs and heats the surface of the neutron 
star (NS), producing a potentially observable X-ray emission component (Arons 1981).  
Two recent developments have introduced important changes to this picture.  First 
was the finding of Muslimov \& Tsygan (1992) that the effect of inertial-frame 
dragging near the NS surface greatly increases the induced electric field 
$E_{\parallel}$ above the PC in space-charge limited flow models 
(Arons \& Scharlemann 1979). The second was the realization that inverse-Compton 
radiation of the primary particles can produce pairs which could potentially 
screen the electric field (Zhang et al. 1997, Harding \& Muslimov 1998; 
hereafter HM98).  

We have investigated the electric field screening and PC heating in the
revised space-charge limited flow (SCLF) model.  In the first paper (Harding
\& Muslimov 2001; Paper I), we have presented our results for screening and
PC heating by CR PFFs, as in Arons (1981), but
using the frame-dragging electric field of Muslimov \& Tsygan (1992).  Paper I
outlined a self-consistent calculation of the PFF height, returning positron
flux and the screening scale length, where the $E_{\parallel}$ and the primary
flux are adjusted
for the change in charge density caused by the returning positrons.  A main
assumption of this calculation was that the screening scale is small compared
to the height of the PFF, the location where the first pairs are produced.
This turns out to be generally true for screening by CR-produced pairs because the pair
cascade multiplicity grows rapidly over small distances due to the strong dependence 
of CR photon energy on particle energy.  Thus, the existence of a CR PFF always
results in PC heating and full screening of $E_{\parallel}$.  
We found that the PC heating luminosity, as a fraction of the spin-down luminosity,
increases with pulsar characteristic age, $\tau = P/2\dot P$, and should be
detectable for pulsars with $\tau \gsim 10^6$ yr.  The most significant heating 
occurs for pulsars near the death line for CR pair production, which is
$\tau \sim 10^7$ yr for normal period pulsars and $\tau \sim 10^8$ yr for millisecond
pulsars.  Our predicted X-ray luminosity due to PC heating is about a factor
of ten higher than the X-ray luminosity predicted by Arons (1981), due to
the increase in accelerating voltage drop resulting from the inclusion of inertial-frame
dragging effects.  We also predicted that older pulsars should have higher
PC surface temperatures from heating.

In this second paper (Paper II), we present results of our investigation of 
electric field screening and PC heating by inverse-Compton scattering
(ICS) radiation PFFs.  We investigate PC heating by ICS produced 
pairs in all pulsars, including those which produce CR pairs.
For the older pulsars that do not produce pairs through CR, PC heating by 
positrons from ICS cascades is especially important.  Our treatment of the 
ICS PFFs follows closely that of HM98, where cyclotron-resonant and non-resonant 
scattering are considered as separate radiation components.  HM98 found that 
the ICS PFFs are located much closer to the NS surface than are the CR PFFs, 
because primaries with Lorentz factor of only $\sim 10^4 - 10^5$ can produce 
pairs via ICS whereas Lorentz factors of at least $\sim 10^7$ are required 
for production of pairs via CR. They also found that the ICS PFFs of positrons 
returning from the upper PFF occur at a significant distance above the surface, 
so it is possible that ICS pairs fronts are unstable if screening of 
$E_{\parallel}$ at the lower PFF occurs.  However, HM98 did not determine 
whether the ICS pairs were capable of screening either at the upper or lower PFFs.  

This paper will attempt to answer those questions with a calculation similar to 
that of Paper I.  In Section II, we describe the acceleration model for this 
calculation.  We use a solution for $E_{\parallel}$ from Poisson's equation 
which imposes an upper boundary condition requiring that $E_{\parallel} = 0$ 
at the location of screening (as in HM98), which differs from that of Paper I 
which used a solution with no upper boundary.  The solution with an upper boundary 
is more appropriate for screening at ICS PFFs which form close to the surface, 
where the screening scale is generally comparable to the PFF height.  This solution 
is required here because pairs from ICS photons may screen $E_{\parallel}$ close 
to the NS surface. In this case the value of $E_{\parallel}$ is suppressed due 
to proximity of the screening to the surface.  Section II also presents a
calculation of ICS PFF height and the location of a new ICS pair death line.  
We find that virtually all known pulsars can produce pairs by either resonant 
ICS (RICS) or non-resonant ICS (NRICS) in a pure dipole field.  
Hibschmann \& Arons (2001) (hereafter HA01) have recently taken a different approach 
to determining pulsar death-lines including ICS produced pairs. They define the 
PFF as the location where the pair multiplicity achieves that required
for complete screening of the $E_{\parallel}$, whereas we define the PFF as the
location where pair production begins (i.e. where the first pairs are produced).
In the case of CR, this distinction is minor since the screening scale is small compared 
to the PFF height, but in the case of ICS the distinction is very important since 
the screening scale is comparable to the PFF height.  HA01 also assume that the pair 
multiplicity required for screening is determined by the difference between the 
actual charge and the Goldreich-Julian charge at a distance from the NS surface 
that is roughly equal to the PC radius, whereas in our calculation the charge density 
required for screening is determined by the charge deficit at the location of creation 
of pairs which can be much smaller. With their more restrictive definition of the PFF, 
their death lines differ 
significantly from ours.  In Section III, we give self-consistent solutions for the 
fraction of positrons returning to the PC and for the screening scale.  We also 
explore the question of whether positrons returning from the upper
PFF can screen $E_{\parallel}$ near the lower PFF close to the NS surface.  The
subject of pulsar pair death lines, i.e. which NSs are capable of pair production,
will be discussed in Section IV.  In Section V, we present our calculations of
PC heating luminosity due to returning positrons, giving both numerical and analytic 
estimates.  Summary and conclusions, as well as a comparison of results from both
Papers I and II will be given in Section VI.

\section{THE ACCELERATION OF PRIMARIES AND ONSET OF PAIR FORMATION}

\subsection{Acceleration model}

In this paper we exploit the same model for charged particle 
acceleration that is described in Paper I. Namely, in the 
acceleration region we use the appropriate solution for the
electric field and potential presented in Paper I. For 
example, for most ICS pair fronts, where the screening 
occurs over the length scale smaller than the PC size, we 
use our formulae (A7)-(A10) of Paper I, whereas in the 
screening region (see \S~3.2 for details) for all cases 
we model the electric field by equation (26) of Paper I.

\subsection{Altitudes of the pair-formation fronts}

\subsubsection{Analytic estimates}

For our analytic estimates we shall use 
the same simplified expressions for the accelerating electric 
field (cf. eqs. [34], [35] of Paper I),
\be
E_{\parallel 6} = {{B_{12}}\over P} \kappa _{0.15} 
\left\{ \begin{array}{ll}
    1.3~z~P^{-1/2} & , \\
    5\cdot 10^{-3}~P^{-1} & ,
\end{array} 
\right.
\label{Epar}
\ee
where the upper expression corresponds to the unsaturated regime 
(rising part of the accelerating field), and the lower - 
to the saturated regime (nearly constant accelerating field). 
Equation (\ref{Epar}) assumes that $\xi = 0.5$ and $\cos \chi \approx 
1$; $E_{\parallel 6} \equiv E_{\parallel }/10^6$ esu, 
$\kappa _{0.15} = \kappa /0.15$, $B_{12} = B_0/10^{12}$ G, 
and $z$ is the altitude in units of stellar radius, where
$\xi$ is the magnetic colatitude in units of the PC half-angle,
$\chi$ is the angle between the magnetic and spin axes, 
$E_{\parallel }$ is the component of the electric field parallel 
to the magnetic field, $B_0$ is the surface value of the magnetic 
field strength, $\kappa $ is the dimensionless general relativistic 
parameter originating from the frame-dragging effect and accounting 
for the stellar compactness and moment of inertia. 

Note that throughout this paper in all practical formulae we 
assume that $P$ is dimensionless value of pulsar spin period measured 
in seconds. Also, in all our analytic estimates based on the 
above formula we shall discriminate between unsaturated and 
saturated regimes of acceleration, with the unsaturated/saturated 
regime occuring in the case where the characteristic altitude 
of pair formation is smaller/larger than the PC radius. The formal 
criteria corresponding to these regimes, e.g. in B-P diagram, 
depend on the radiation mechanism for pair producing photons 
and can be derived from the conditions $\zeta _{\ast } \lsim 1$ 
and $\zeta _{\ast } \gsim 1$ for unsaturated and saturated regimes, 
respectively, where $\zeta _{\ast }$ is the characteristic altitude 
of screening scaled by the PC radius (see 
eqs. [\ref{zeta*CR}]-[\ref{zeta*NR}] below). These criteria 
translate into 
\be
P \lsim P_{\ast }~~~~~{\it Unsaturated~regime}
\ee
and 
\be
P \gsim P_{\ast }~~~~~{\it Saturated~regime},
\ee
where  $P_{\ast }$ is defined as 

\noindent{\it Curvature radiation}
\be
P_{\ast }^{(CR)} = 0.1~B_{12}^{4/9}.
\ee
Before we present $P_{\ast }$ for the ICS case, note that 
for the physical conditions we discuss 
in this paper it is convenient (especially in our analytic 
calculations) to treat the ICS photons as generated via 
two different regimes of scattering: resonant (R) and non-resonant 
extreme Klein-Nishina scattering (NR). In reality, the 
spectrum of ICS photons from an electron of Lorentz factor 
$\gamma $, a subset of which produce pairs, may be produced 
by scatterings in both the R and NR regimes. However, the 
first pairs which mark the PFF location will come from one 
of the two regimes. Thus, in our analytic calculations, to avoid 
unnecessary complexity, we shall differentiate between R  
and NR regimes. Now let us present the expressions 
for $P_{\ast }$ corresponding to these regimes

\noindent{\it Resonant ICS}
\be
P_{\ast }^{(R)} = 0.1~B_{12}^{6/7},
\ee
\noindent{\it Non-resonant ICS} 
\be
P_{\ast }^{(NR)} = 0.4~B_{12}^{4/7}.
\ee

In this Section we shall explore the altitudes of PFFs produced 
by the ICS photons. To estimate the altitude of the PFF above the stellar 
surface we can use, as in our previous papers 
(see HM98, eq. [1]; and Paper I, eq. [37]), 
the following expression 
\be
S_0 = {\rm min} [ S_a (\gamma _{\rm min}) + S_p (\varepsilon _{\rm min})], 
\label{S0}
\ee
where $S_a(\gamma _{\rm min})$ is the acceleration length 
that is required for an electron to produce a photon of energy 
$\varepsilon _{\rm min}$, and $S_p(\varepsilon _{\rm min})$ is the 
photon pair-attenuation length. 

\noindent{\it Resonant ICS}

In the R regime of ICS the characteristic energy of a scattered 
photon is 
\be
\varepsilon \sim 2 \gamma B' ,
\label{}
\ee
where $B' \equiv B/B_{cr} $ ($< 1$) is the local value of the magnetic 
field strength in units of critical field strength, 
$B_{cr} = 4.41\times 10^{13}$ G. 

In this paper in our analytic estimates we assume that $B\approx B_0$, 
which is justified for all cases except for the case of millisecond pulsars. 
By substituting $\varepsilon $ into the expression for $S_p$ (see HM98 or 
Paper I for details) we get 
\be
S_p^{(R)}  = C_{\gamma }^{(R)}/\gamma ,
\ee
where $C_{\gamma }^{(R)} = 2.2\times 10^{10} P^{1/2}/B_{12}^2~cm$ . 

As far as the acceleration length $S_a$ is concerned, it is independent 
of the radiation mechanism (provided that the radiation losses are negligible 
during acceleration), and the formulae of Paper I (see eqs. [38]) are still 
applicable. After substituting expressions for $S_a$ and $S_p^{(R)}$ 
into (\ref{S0}) we find that $S_0$ is minimized at 
\be
\gamma_{\rm min}^{(R)} = 10^6 \left\{ \begin{array}{ll}
    0.9~P^{-1/6}B_{12}^{-1} & {\rm if}\: P\lsim P_{\ast }^{(R)}, \\
    0.2~P^{-3/4}B_{12}^{-1/2} & {\rm if}~P\gsim P_{\ast }^{(R)}.
\end{array} 
\right.
\label{GamR}
\ee

Now we can evaluate expression (\ref{S0}) at $\gamma _{\rm min}$ 
to calculate the dimensionless altitude (scaled by a NS radius)
of the PFF
\be 
z_0^{(R)} \equiv S_0^{(R)}(\gamma _{\rm min}^{(R)})/R = 10 ^{-2}
\left\{ \begin{array}{ll}
    7~P^{2/3}B_{12}^{-1} & {\rm if}\: P\lsim P_{\ast }^{(R)}, \\
    17~P^{5/4} B_{12}^{-3/2}& {\rm if}~P\gsim P_{\ast }^{(R)}.
\end{array} 
\right.
\label{z0R}
\ee

\noindent{\it Non-resonant ICS}

In the NR regime we can use the extreme Klein-Nishina 
formula, and write for the energy of the scattered photon 
\be
\varepsilon \sim \gamma .
\ee
The corresponding photon pair-attenuation length is 
\be
S_p^{(NR)} = C_{\gamma }^{(NR)}/\gamma ,
\ee
where $C_{\gamma }^{(NR)} = 10^9 P^{1/2}/B_{12}~cm$.
The expressions for $\gamma _{\rm min}^{(NR)}$ and $z_0^{(NR)}$ then 
read
\be
\gamma_{\rm min}^{(NR)} = 10^5 \left\{ \begin{array}{ll}
    P^{-1/6}B_{12}^{-1/3} & {\rm if}\: P\lsim P_{\ast }^{(NR)}, \\
    0.6~P^{-3/4} & {\rm if}~P\gsim P_{\ast }^{(NR)},
\end{array} 
\right.
\label{GamNR}
\ee
and
\be 
z_0^{(NR)} = 10 ^{-2}
\left\{ \begin{array}{ll}
    3~(P/B_{12})^{2/3} & {\rm if}\: P\lsim P_{\ast }^{(NR)}, \\
    4~P^{5/4} / B_{12} & {\rm if}~P\gsim P_{\ast }^{(NR)}.
\end{array} 
\right.
\label{z0NR}
\ee
The unsaturated regime of formula (\ref{GamNR}) agrees within a factor 
of $\sim $ 2 with our numerical calculations for the ms and middle-aged 
pulsars.  

\subsubsection{Numerical calculations}

The location of the PFF for the different pair-producing radiation processes 
is more accurately determined using numerical calculations of the minimum 
height at which the first pairs are produced. Here we incorporate full spatial 
dependence (in $r$ and $\theta$) of quantities such as the magnetic field 
and radius of curvature in computing the altitude of the PFF.  However, as we will 
discuss in Section 3.2, we use a one-dimensional model for the pair
dynamics to treat the screening of the electric field about the PFF.
The numerical calculation of the altitude of the PFF follows closely that of HM98, 
but with one notable exception.  As discussed above,
we use the solutions for $E_{\parallel}$ with an upper boundary 
to compute the ICS PFFs (as in HM98) for normal pulsars, whereas 
we use the solutions for $E_{\parallel}$ with no upper boundary (but which saturates
at infinity) to compute the CR PFFs (as in Paper I) and the ICS PFFs for ms pulsars.  
Otherwise, we have used the same expressions for the energy of pair 
producing photons: equation (29) of HM98 for CR photons and equation (43) of HM98 
for ICS photons.  We have also made the assumption (as in HM98) that the primary
electrons travel along the magnetic axis in computing the ICS energy loss and
scattering rate.  Unlike in the analytic estimates presented, we do not 
separate the RICS from the NRICS photons but take a weighted average for 
the ICS photon energy:
\be
\langle \epsilon ^{(ICS)} \rangle = { {\langle \epsilon ^{(R)} \rangle\dot\gamma ^{(R)}
+ \langle \epsilon ^{(NR)} \rangle\dot\gamma ^{(NR)}}\over {\dot\gamma ^{(R)} + 
\dot\gamma ^{(NR)}} }.
\ee
Calculations of $z_0$ as a function of the colatitude $\xi$ for four different
sets of pulsar parameters are shown in Figure 5.  Comparison of Figures 5a and 5b
confirms that the value of $z_0$ is independent of PC temperature, since 
RICS photons have average energies of $2\gamma B'$ and NRICS occurs primarily in
the extreme Klein-Nishina limit, where the average scattered photons energy 
is $\gamma$.

With numerical calculation of the PFF altitude, we are able to determine the pulsar
parameter space in which the formation of a PFF by the different photon processes is 
possible (i.e. where $S_p(\epsilon_{\rm min}) < \infty$).  
Figure 1 shows the CR and ICS PFF parameter space as a function of 
pulsar period $P$ and surface field strength $B_0$, as determined by the dipole formula 
$B_0 = 6.4 \times 10^{19}\,{\rm G}\,(\dot P P)^{1/2}$.  We have also plotted 
observed pulsars from the ATNF catalog (http://www.atnf.csiro.au/~pulsar/), 
which includes pulsars from the 
Parkes Multibeam Survey (Manchester et al. 2001)
having measured $\dot P$. For the $B_0$ and $P$ values above the lines in Fig. 1 for the
different processes, we were able to find numerical 
solutions to equation (\ref{S0}).  Pulsars below
the lines are therefore not capable of producing pairs from that process.  We find 
that the majority of pulsars cannot produce pairs through CR.  This result has 
been well known for some time (Arons \& Scharlemann 1979, Arons 1998, 
Zhang et al. 2000).  We find that virtually all known pulsars, with the exception of 
only a few ms pulsars, are capable of producing pairs via ICS. For the majority of 
pulsars, those with $B_0 \lsim 0.1 B_{\rm cr}$, the pairs are produced from NRICS 
photons.  We have plotted ICS PFF boundaries for three different 
temperatures and, as expected, there is only a small dependence on PC  
temperature.  The slight variation due to PC temperature occurs for higher field
pulsars where scattering is not in the extreme Klein-Nishina regime and the photon 
energies become somewhat temperature dependent.  The failure of the lowest-field 
ms pulsars to produce ICS pairs in our calculation may be due to the failure of our 
$E_{\parallel}$ solution, obtained in the small-angle approximation, to accurately
model the accelerating field in the ms pulsars which have large PCs with 
$\theta_{pc} \sim 0.3$; or to our use of a canonical NS model in the case of ms 
pulsars which may have undergone significant mass accretion in Low-Mass Binary 
systems. We are 
currently exploring these effects to correct this shortcoming.  Also, we shall 
examine the possibility of photon-photon pair formation in ms pulsars. We will 
discuss the broader implication of pair death lines in Section 4.

\section{Returning positron fraction and electric field 
screening by pairs}

\subsection{Analytic estimates}

To estimate the maximum fractional density of positrons 
returning from the upper PFFs we can use equation (33) of Paper I 
and the expressions for $z_0$ derived in \S~2.2 . We must note, 
however, that in Paper I (where we investigated the CR case alone) 
in our analytic calculations of the fraction of returning positrons, 
we justifiably assumed that the screening occurs within the 
length scale much less than $z_0$. This is not the case for 
ICS, for which the screening scale is 
determined by the scale of growth of energy of pair-producing 
photons. For the ICS photons the energy of pair-producing photons 
$\sim \gamma $, so that the scale of growth of $\gamma $ governs 
the screening scale $\Delta_s$, i.e. 
\be 
\Delta _s \sim \left[ {\gamma \over {d\gamma /dz}} \right] _{z=z_0} 
\approx  
\left\{ \begin{array}{ll}
    0.5~z_0 & {\rm if}\: P\lsim P_{\ast }^{(R,NR)}, \\
    z_0 & {\rm if}~P\gsim P_{\ast }^{(R,NR)}.
\end{array} 
\right.
\label{Deltas}
\ee
Thus, in the case of ICS the screening effectively occurs over a
more extended region. This means that even within the screening 
region the primary electrons keep accelerating and generating 
pair-producing photons. Since the pair-production rate per primary 
electron in the case of the ICS photons is not as high as 
in the case of CR photons, the assumption that the altitude of 
the screening is $z_0$ would result in a significant underestimation of 
the returning positron fraction at the ICS-controlled pair-fronts. To come 
up with better estimates we should add $\Delta_s$ to the 
calculated values of $z_0$, or, according to formula (\ref{Deltas}),
multiply $z_0$ by a factor 1.5 and 2, for the unsaturated and 
saturated regimes, respectively.

Evaluating expressions for $\rho _+/\rho _{GJ}$ (here $\rho _+$ is 
the charge density of returning positrons, and $\rho _{GJ}$ is the Goldreich-Julian 
charge density; see also eq. [33] of Paper I) at $z_{\ast }=z_0+\Delta _s$, we get 

\noindent{\it Resonant ICS}

\be 
\left({{\rho _+}\over {\rho _{GJ}}}\right)^{(R)} = 
10^{-2} 
\left\{ \begin{array}{ll}
    2~P^{2/3}/B_{12} & {\rm if}\: P\lsim P_{\ast }^{(R)}, \\
    5~P^{5/4} / B_{12}^{3/2} & {\rm if}~P\gsim P_{\ast }^{(R)},
\end{array} 
\right.
\label{frac_R}
\ee
and

\noindent{\it Non-resonant ICS}

\be 
\left({{\rho _+}\over {\rho _{GJ}}}\right)^{(NR)} = 
10^{-3} 
\left\{ \begin{array}{ll}
    7~(P/B_{12})^{2/3} & {\rm if}\: P\lsim P_{\ast }^{(NR)}, \\
    9~P^{5/4} /B_{12} & {\rm if}~P\gsim P_{\ast }^{(NR)}.
\end{array} 
\right.
\label{frac_NR}
\ee
In their paper HA01 assume that the fraction of returning 
positrons (see \S~5.3) is of order of 
$\kappa r_{pc}/R$, where $r_{pc} (\approx 10^4/ \sqrt{P}$ cm) is 
the PC radius, and $R$ is the NS radius ($= 10^6~cm$). 
In contrast, in our calculations the 
fractional density of returning positrons is proportional to 
the dimensionless screening altitude (see Paper I), and 
can be roughly estimated as $1.5 [\kappa /(1-\kappa )] z_0$. 
Thus, the ratio of our value for the density of returning 
positrons to that of HA01 is of order of $S_0/r_{pc}$, which 
for the observed pulsar parameters may range from $\sim 0.1$
to $\sim 20$.  More specifically, for short-period pulsars 
our calculated fractions of returning positrons are up to 
a factor of $\sim 10$ less than those of HA01, while for 
relatively long-period pulsars they are larger by about the same 
factor. Besides this quantitative difference, there is 
a significant difference in the dependence of our fractional 
densities on $B$ and $P$.  Note that in reality, the screening 
scale and thus also $\rho_+/\rho_{GJ}$ is dependent on the PC temperature, 
as will be shown in the numerical calculations.  We cannot model this 
dependence analytically since we have assumed that the $\Delta_s$ is 
a constant multiple of $z_0$ for all pulsars.

\subsection{Numerical calculations}

Our numerical calculation of the returning positron fraction and scale
length of the ICS pair screening follows that described in Paper I
for CR pair screening.  As detailed in that paper, we first compute the
pair source function in energy and altitude above the PFF and then compute
the dynamical response of the pairs to the $E_{\parallel}$ above the PFF
by solving the continuity and energy equations of the pairs to obtain
the charge density.  The $E_{\parallel}$ above the PFF is parameterized
as an exponential with scale height $\Delta_s$, which is readjusted at
each iteration to equal a multiple of the height at which the computed
charge density from the pairs equals the difference between the primary
beam charge density and Goldreich-Julian charge density.  In
computing the source function of pairs from ICS photons, we consider a
hybrid spectrum of RICS and NRICS photons from the primary particle, of Lorentz 
factor $\gamma$, at a given altitude of its acceleration.  In the ICS photon
spectrum, photons with energies $\epsilon < 2\gamma B'$ are produced primarily
by RICS.  Since the RICS spectrum from a single electron cuts off sharply 
for photon energies $\epsilon \simeq 2\gamma B'$, photons above this energy,
up to a maximum energy of $\epsilon = \gamma$ will be produced by NRICS.
We therefore describe the ICS photon spectrum for $\epsilon < 2\gamma B'$
by a RICS spectrum based on an expression given by Dermer (1990).  Assuming
delta-function distributions of electrons of Lorentz factor $\gamma$, and
thermal photons of energy $\epsilon_0 = 2.7~kT/mc^2$ and incident angles 
$\mu_- < \mu < \mu_+$ in the lab frame, Dermer's (1990) equation (10) for 
the distribution of scattered photons with energy $\epsilon_s$, integrated
over scattered angles, $-1 < \mu_s < 1$, becomes
\be \label{N^R}
{N^{R}(\epsilon_s)\over dtd\epsilon_s} = {c\sigma_T \epsilon_B\over 2(\mu_+ - \mu_-)}
{1\over \beta^2\gamma^3}{n_0\over\epsilon_0^2}[J_0 + J_1 + {1\over 2}(J_2-J_1)] ,
\ee
where $J_n = I_n(\mu_+) - I_n(\mu_-)$,
\begin{eqnarray} \label{u}
u_+ = \gamma\epsilon_B^{-1}\, {\rm max}[\epsilon_s/2\gamma^2, 
\epsilon_0(1-\beta\mu_+] ,               &   \nonumber \\
u_- = \gamma\epsilon_B^{-1}\, {\rm min}[2\epsilon_s, 
\epsilon_0(1-\beta\mu_-)] ,              &
\end{eqnarray}
and
\begin{eqnarray}
I_0(u) & = & {3\over (4\beta\gamma\epsilon_B)^4}\,\left[ {-\epsilon_0^2\epsilon_s^2
\over 3u^3} + {\gamma\epsilon_0\epsilon_s\epsilon_B(\epsilon_0+\epsilon_s)\over
u^2} \right. \nonumber \\
& - & \left. {\epsilon_B(\epsilon_0^2+\epsilon_s^2+4\gamma^2\epsilon_0\epsilon_s)\over u}
- 2\gamma\epsilon_B^2(\epsilon_0+\epsilon_s)\ln u + \epsilon_B^4 u \right] , 
\end{eqnarray}
\be
I_1(u) = u - (u+1)^{-1} - 2\ln(u+1) ,
\ee
\be
I_2(u) = u + \ln[(u-1)^2 + a^2] + (a^{-1} + a)\tan^{-1}\left[{(u-1)\over a}\right] ,
\ee
where $\sigma _T$ is the Thompson cross section, $\beta $ is the electron velocity 
in units of the velocity of light,  
$\epsilon_B = B/B_{\rm cr}$ is the cyclotron energy in units of $mc^2$,
$a = 2\alpha_f \epsilon_B/3$, $\alpha_f$ is the fine-structure constant
and $n_0 = 20\,{\rm cm^{-3}}\,T^3$ is the density of thermal photons of energy 
$\epsilon_0$. 
Each scattered photon is assumed to be emitted along the direction of the
particle momentum (i.e. along the local magnetic field).  Although the distribution
of scattered photon angles depends on energy (see Fig. 3 of Dermer 1990), the
photons at pair-producting energies scattered by relativistic electrons are mostly 
emitted along the field direction.  Since we will be treating scattering of both 
upward-moving electrons and downward-moving positrons, we have
\begin{eqnarray}  \label{mupm}
\mbox{for~electrons:}&&\mu_- = \mu_c \nonumber \\ 
&&\mu_+ = 1 \\
\mbox{for~positrons:}&&\mu_- = -1 \nonumber \\
&&\mu_+ = -\mu_c 
\end{eqnarray}
in equation (\ref{u}), where
\be
\mu_c = \cos\theta_c = {h\over \sqrt{h^2 + r_{\rm t}^2}}
\ee
is the largest scattering angle at height $h$ above the NS surface,  
and $r_{\rm t} = r_{\rm pc} \ll R$ is the radius of the hot PC. 

The ICS photon spectrum for $\epsilon > 2\gamma B'$ is produced by NRICS.
We use a spectrum for relativistic (but non-magnetic) scattering
which is based on an expression of Jones (1968), derived originally to
apply to scattering of an isotropic distribution of photons by a relativistic
electron, which we have modified to apply to a semi-isotropic distribution
of thermal photons
\be  \label{N_NR}
{N^{NR}(\epsilon_s)\over dtd\epsilon_s} = {n_0\pi r_0^2 c\lambda _{\pm}\over \gamma^2\epsilon_0}
\left[2q\ln q + (1+2q)(1-q) + {(q\Gamma_0)^2\over (1+q\Gamma_0)}{(1-q)\over 2}
\right] ,
\ee
where
\be
q = {E_s\over \Gamma_0(1-E_s)}, ~~~~E_s = {\epsilon_s\over \gamma}, 
~~~\Gamma_0 = 2\gamma\epsilon_0\lambda _{\pm} .
\ee
Here $r_0$ is the classical electron radius. For upward-moving electrons, 
$\lambda  _- = (1 - \beta\mu_c)$ and for downward-moving positrons, $\lambda _+ = 2$.

Our method for calculating the pair source function from ICS photons is 
similar to that used in Paper I for CR photons.  However, unlike in the case of CR, 
higher generations of pairs from synchrotron radiation of first-generation pairs are 
important in the screening, since the attenuation length for the sychrotron photons
is less than the ICS screening scale.  As in Paper I, the contribution to the
pair source function from first generation pairs is computed by dividing the 
ICS spectrum radiated by the primary particle at each step along its path into
discrete energy intervals.  A representative photon from each ICS energy bin
is propagated through the local field to determine whether it
escapes or produces a pair, in which case the location of the pair is recorded 
(for details of such a calculation, see Paper I and Harding et al. 1997).  The
pairs for each ICS spectral energy interval of width $\Delta\epsilon_s$ 
are then weighted in the source function by the ``number" of RICS or NRICS photons,
$n^{\rm R,NR}$, represented by the test photon in that energy bin, estimated by
integrating the photon spectrum over the bin
\be
n^{\rm R,NR}(\epsilon_s) = {\Delta s\over c}\,\int_{\epsilon_s-{\Delta\epsilon_s\over 2}}
^{\epsilon_s+{\Delta\epsilon_s\over 2}}\,{N^{R,NR}(\epsilon_s')\over dtd\epsilon_s'}
d\epsilon_s',
\ee
where $\Delta s$ is the size of the particle path length.  

The contribution of higher generations of pairs to the source spectrum is 
computed by simulating a pair cascade in the method described by 
Baring \& Harding (2001), except that we do not include the possibility 
of photon splitting in the present calculation. By limiting the surface 
magnetic field we consider to $B_0 < B_{\rm cr}$, we can safely neglect 
splitting.  As described in Baring \& Harding (2001), the created
pairs from the first generation make transitions between Landau states to radiate
synchrotron/cyclotron photons until they reach the ground Landau state.  Each
synchrotron/cyclotron photon is individually traced through the local magnetic field
until it either escapes or creates a pair, which radiates another generation of 
photons.  A recursive routine is called upon the radiation of each photon,
so that a large number of pair generations may be simulated.  

One notable difference was introduced in the present cascade simulation in order to treat
the synchrotron radiation from ms pulsars.  The low surface fields of the ms pulsars,
combined with the photon energies $\epsilon \sim 10^6 - 10^7$ required
to produce pairs, give very large Landau states for the created pairs, making it
impossible to treat the synchrotron/cyclotron photon emission discretely.  For
local fields $B < 0.01B_{\rm cr}$, we treated the synchrotron radiation from the pairs
by the same method as for the ICS radiation from the primaries, i.e. 
as a continuous spectrum.
To describe the total spectrum radiated by a particle as it decays from a
high initial Landau state (i.e. large initial pitch angle) to the ground state
(zero pitch angle) we use the calculated synchrotron spectrum of Tademaru (1973)
\be
N_s(\epsilon)d\epsilon = {1\over 2}(B'\sin\psi_0)^{-1/2}\epsilon^{-3/2}d\epsilon ,
\ee
where $\psi_0$ is the initial pitch angle of the particle and $B' = B/B_{\rm cr}$
is the local magnetic field strength.  Surprisingly, for the ms pulsars we are
required to limit the upper end of the synchrotron spectrum at an energy equal to 
the particle Lorentz factor in order to avoid violating energy conservation.
This is a well-known problem in high magnetic fields (e.g. Harding \& Preece 1987), 
but in this case the very high Lorentz factors of the pairs give a synchrotron critical
energy $\epsilon_{\rm cr} = (3/2)\gamma^2 B'\sin\psi_0 > \gamma$.  Imposing the
energy conservation condition
limits the number of pair generations.  The discrete QED treatment of the synchrotron
emission in high magnetic fields (described above) is able to treat the effect of 
pair production in the ground Landau state, which also severely limits the number of pair 
generations in fields $B \gsim 0.2B_{\rm cr}$ (Baring \& Harding 2001).

\subsubsection{Screening effect at the upper pair front}

Since the ICS photon spectrum has a much weaker dependence on the particle
Lorentz factor than the CR photon spectrum, the ICS pair source function grows 
much more slowly with height above the PFF.  In the case of NRICS, the photon 
emission rate decreases as $\sim 1/\gamma$, while the RICS emission 
rate first increases sharply at low $\gamma$ as the soft photons come into 
resonance in the particle rest frame, and then decreases as $1/\gamma^2$, 
whereas in the case of CR the photon emission rate increases as $\gamma$.  
Even though the characteristic ICS photon energies are increasing with 
$\gamma$, the stronger decrease in photon number with height as the particle 
accelerates, combined with the decrease in soft photon density and angle with 
altitude, makes screening by pairs produced by ICS photons much more difficult 
and in some cases 
impossible. In nearly all cases, the screening scale lengths are comparable to 
the particle acceleration length, requiring different treatment of the self-consistent
screening calculation from what was used in Paper I for CR screening.  
The onset of pair production and possible screening close to the NS
surface couples the entire $E_{\parallel}$ solution to the screening scale.
We therefore must correct the accelerating $E_{\parallel}$ as well as the 
$E_{\parallel}$ in the screening region above the PFF on every iteration of
the screening scale length.  This is done by adjusting the upper boundary of the
$E_{\parallel}$ solution (eqs. [A7] - [A10] of Paper I) to $z_0 + \Delta_s$, 
so that the electric potential distribution below the pair front is accurate.  We still 
assume the exponential form with scale height $\Delta_s$ (eq. [26] of Paper I) for 
$E_{\parallel}$ above the PFF.  But unlike our calculation of CR pair screening, 
the location of the PFF will change and adjust on every interation of $\Delta_s$. 
The exponentional function $E_{\parallel }^{sc}(z \gsim z_0) =
E_{\parallel }^{acc}(z_0) \exp [-(z-z_0)/\Delta _s]$
describing the $E_{\parallel}$ above the PFF
is also renormalized at the PFF at each iteration so that the value
of $E_{\parallel }^{acc}(z_0)$ is equal to the maximum 
value of the $E_{\parallel}$ solution between $z = 0$ and $z = 
z_0 + \Delta_s$.  We also must take into account the increase in the charge 
deficit in the screening 
region, which will also cause ICS screening to be more difficult. 

Figure 2 shows an example of an ICS pair source function, integrated over energy,
so that the growth of pairs as a function of distance $x = z - z_0$ 
above the PFF is displayed.
The first generation pairs (those produced by primary electrons) is almost flat
before beginning to decrease.  Including pairs from higher cascade generations
produces an initially sharp increase in number of pairs vs. height, but the pair
number quickly levels off and also eventually decreases.  The ICS pair source
function is a sharp contrast to the CR pair source function shown in Figure 1
of Paper I, which increases nearly exponentially with height above the PFF.
This raises the interesting question of whether ICS pairs are capable of screening
$E_{\parallel}$ to all altitudes, even if they are able to initially screen the
local $E_{\parallel}$.  Figure 3 shows an example of a self-consistent solution
for the charge density $\rho(x)/\rho_{\rm GJ}$ due to polarization of pairs and 
$\Delta _s$ in the screening 
region.  The charge density initially increases faster than the charge deficit,
which is increasing approximately as $x$.  In this case, screening of 
$E_{\parallel}$ is achieved locally, until the pair source function begins to
decline with $x$, causing the charge density due to pairs to decrease.  
As the charge density
due to pairs decreases while the charge deficit is increasing, screening cannot be
maintained and the $E_{\parallel}$ will again begin to grow.  We cannot model
this effect in the present calculation, which would require the solution of
Poisson's equation throughout the screening region.  However, it is clear that
ICS screening cannot short-out $E_{\parallel}$ and produce a limit to particle 
accleration at low altitudes.   ICS screening, even if locally complete, will
only slow the particle acceleration until the ICS pair density falls off and 
beyond this point the particles will resume their acceleration until they can
produce CR pairs, which will then screen the $E_{\parallel}$ at higher altitudes.

For many pulsars, however, ICS cannot even achieve locally complete screening.
Figure 4 shows computed boundaries of ICS local screening for different assumed
PC temperatures, $T_6 \equiv T/10^6$ K, as a function of pulsar surface 
magnetic field and period.  Pulsars above and to the left of these boundaries
can achieve local screening of $E_{\parallel}$ at a colatitude of $\xi = 0.7$,
while pulsars below and to the right of the boundaries cannot.  Although the PFF 
boundaries were nearly independent of $T_6$, the screening boundaries are very 
sensitive to the temperature of the soft photons.  For low $T_6$, only very 
high-field pulsars, those whose pairs source functions are dominated by RICS, 
can screen with ICS photons.  As $T_6$ increases, the boundaries move
down, but are limited in their movement to the right as they approach the ICS PFF
boundaries (Figure 1).  There is also some dependence of the screening boundaries on $\xi$,
in that the boundary for a given $T_6$ will be higher at a lower value of $\xi$.
This means that screening is more difficult in the inner parts of the PC, so that
some pulsars may have local ICS screening only in the outer part of their PC.

Even though complete ICS screening is not achieved for all pulsars, there will 
still be positrons which turn around and return to heat the PC.  In fact this heating
can be substantial since in the absence of complete screening which shuts of
the acceleration, most of the positrons from ICS pairs will return to the PC. 
Figure 5 shows self-consistent calculations of the returning positron fraction,
as well as the PFF altitude $z_0$ which we have discussed in Section (2.2), as a
function of $\xi $.  Figures 5a and 5b, which have the same $P$ and $B$ values but
temperatures $T_6 = 0.5$ and $T_6 = 3.0$ respectively, illustrate that although 
there is very little difference in the PFF altitude with a large increase in PC 
temperature there is a large increase in returning positron fraction.  The
dependence of $\rho^+/\rho_{\rm GJ}$ on $T_6$ is not reflected in the analytic 
estimates (eqs. [\ref{frac_R}], [\ref{frac_NR}]) which are based only on a
constant multiple of the charge 
deficit at the PFF altitude, proportional only to $z_0$.  
The size of $\rho^+/\rho_{\rm GJ}$ is actually dependent on the number of 
pairs produced above $z_0$ which is not dependent on the charge deficit if 
screening is not achieved.  Comparison of Figures 5a and 5c shows that pulsars 
with larger periods have relatively larger $z_0$ and also larger fractions of 
returning positrons.  This is due to the fact that longer period pulsars have 
more difficulty producing pairs due to the smaller PC size and therefore larger 
radii of curvature of the last open field line. As illustrated in Figure 5d, 
ms pulsars have very large $z_0$ and large fractions of returning positrons,
but smaller than expected from the analytic solutions, because they do not
produce enough pairs to screen the field. 

Figure 6 shows the dependence of returning positron fraction on PC 
temperature $T_6$ and surface field strength.  The steep dependence of 
$\rho^+/\rho_{\rm GJ}$ on $T_6$ at low temperatures $T_6 < 1.0$ is due to
the increase in numbers of high energy ICS photons that produce pairs.  For
$T_6 > 1.0$, in the case of normal pulsars there is local screening at these 
field strengths, which limits
the number of returning positrons to be proportional to $(z_0 + \Delta_s)$ 
rather than to
$T_6$. The screening therefore produces a saturation in the PC heating, which
will be seen in the calculation of PC heating luminosities to be presented in
Section V.

\subsubsection{Screening at lower pair fronts}

As discussed in HM98, the positrons returning from the upper PFF will be accelerated
through the same voltage drop as the primary electrons and thus will reach the same
high energies as the primaries that produced them.  They will therefore initiate
pair cascades as they approach the NS surface.  HM98 speculated that the pairs
from these downward moving cascades may screen the $E_{\parallel}$ above the
surface since they found that the PFFs of the positrons are located at significant
altitudes above the surface, due to asymmetry in the ICS between upward moving
and downward moving particles.  Now that we can calculate the fraction of positrons
returning from the upper ICS PFF, we can investigate the question of screening at the
lower PFF.  We have simulated cascades below the PFFs of test positrons returning 
from the upper PFF.  The source functions of pairs from the downward cascades and
the charge density are computed in the same way as at the upper PFF, and the
screening scale height is determined self-consistently.  But there are several
critical differences between screening at upper and lower PFFs.  First, the flux of 
returning positrons from the ICS PFF is small compared to the primary flux and the charge
density must be weighted with the value $\rho^+/\rho_{\rm GJ}$.  Second, the value of
$E_{\parallel}$ is very small near the NS surface, making it harder to turn around
the electrons from the pairs.  Third, the charge deficit is much smaller than at
the upper PFF and decreasing toward the stellar surface, being proportional to the
altitude of the returning positron pair front.
This guarantees that there will be a point above the NS surface where 
$\rho(x)/\rho_{\rm GJ}$ will equal the charge deficit. 

We have computed pair source functions from cascades of downward-moving positrons
and the resulting charge densities below the lower PFF for a range of pulsar
parameters.  We find that for nearly all pulsars, the charge densities do not
reach values high enough to screen $E_{\parallel}$ significantly above the NS surface.
For pulsars having surface fields $B \lsim 0.1 B_{\rm cr}$ and periods $P < 0.5$, 
pair multiplicities from the near-surface cascades are not high enough to balance
the low fraction of returning positrons, primarily because most pairs are produced
in the ground Landau state in the high near-surface fields.  For pulsars having
lower surface fields and longer periods, screening can occur for some cases just above 
the surface at the highest PC temperatures.  These pulsars are near the ICS screening
boundary (Figure 4).  Therefore, according to our calculations it seems that ICS
PFFs will be stable except possibly near the screening boundaries. 

\section{Energetics of primary beam and pulsar death line}

\subsection{Luminosity of primary beam}

Before we discuss the pulsar death line, it is 
instructive to calculate the luminosity of the primary electron beam 
at the PFF. 

The luminosity of the primary beam as a function of $z$ ($\ll 1$) can 
be written as (see eq. [73] of MH97, and cf. eqs. [61] of Paper I)
\be \label{Lprim_gen}
L_{\rm prim} = \alpha c \int _{S(z)} \rho (z, \xi) \Phi (z,\xi)dS, 
\ee
where $\alpha = \sqrt {1-r_g/R}$ ($r_g$ is the gravitational 
radius of a NS of radius R), $\rho $ is the charge density of the
electron beam, $\Phi $ is the electrostatic potential, $dS = r_{pc}^2 
d\Omega _{\xi }$, and $d\Omega _{\xi } = \xi d\xi d \phi $ is an 
element of solid angle in the PC region (see also Paper I for 
details). Here the integration is over the area of a sphere cut 
by the polar flux tube at altitude $z$.

Similar to our derivation of formulae (63)-(65) of Paper I, we  
can write for the luminosity of the primary beam 
\be
L_{\rm prim} = f_{\rm prim} {\dot E}_{\rm rot} ,
\label{Lprim}
\ee
where ${\dot E}_{\rm rot}$ is the pulsar spin-down luminosity 
($= \Omega ^4B_0^2R^6/6c^3f^2(1)$, where $B_0/f(1)$ is the 
surface value of the magnetic field strength corrected for the 
general-relativistic red shift; see eq. [66] of Paper I for details), and 
$f_{\rm prim}$ is the efficiency of converting spin-down luminosity 
into the luminosity of the primary beam. Thus, by comparing the 
above two equations, we can get the following expression for 
$f_{\rm prim}$ 
\be
f_{\rm prim} = 6\cdot 10^{-3} P^{-1/2} \zeta 
    \left\{ \begin{array}{ll}
    \zeta & {\rm if}\: P\lsim P_{\ast }^{(CR,R,NR)}, \\
    0.6 & {\rm if}~P\gsim P_{\ast }^{(CR,R,NR)},
\end{array} 
\right.
\label{fprim}
\ee
where $\zeta $ is the altitude scaled by the PC radius, 
$r_{pc}$. In the derivation of formula (\ref{fprim}) we assumed 
$\kappa = 0.15$ and $\cos \chi \approx 1$. 

\noindent
Now we should evaluate expression (\ref{fprim}) at the altitude beyond the 
PFF where the electric field is screened, or at 
$\zeta _{\ast } = \zeta _0 + \Delta _s R/r_{pc}$. For the CR case 
$\zeta _{\ast } \approx \zeta _0$, while for the ICS pair fronts 
(for both R and NR regimes; see also 
eq. [\ref{Deltas}])
\be
\zeta _{\ast } \approx 
\left\{ \begin{array}{ll}
    1.5 \zeta _0 & {\rm if}\: P\lsim P_{\ast }^{(R,NR)}, \\
    2 \zeta _0 & {\rm if}~P\gsim P_{\ast }^{(R,NR)} .
\end{array} 
\right.
\ee
After substituting $\zeta _0$ we get the following explicit 
expressions for $\zeta _{\ast }$ 

\noindent{\it Curvature radiation}

\be  
\zeta _{\ast }^{(CR)} = 
\left\{ \begin{array}{ll}
    30~P^{9/7}/B_{12}^{4/7} = 2~B_{12}^{5/7}\tau _6^{9/14} & {\rm if}\: P\lsim P_{\ast }^{(CR)}, \\
    430~P^{9/4} / B_{12} = 4~B_{12}^{5/4}\tau _6^{9/8} & {\rm if}~P\gsim P_{\ast }^{(CR)},
\end{array} 
\right.
\label{zeta*CR}
\ee

\noindent{\it Resonant ICS}

\be 
\zeta _{\ast }^{(R)} = 
\left\{ \begin{array}{ll}
    10~P^{7/6}/B_{12} = B_{12}^{1/6}\tau _6^{7/12} & {\rm if}\: P\lsim P_{\ast }^{(R)}, \\
    30~P^{7/4} /B_{12}^{3/2} = B_{12}^{1/4}\tau _6^{7/8} & {\rm if}~P\gsim P_{\ast }^{(R)},
\end{array} 
\right.
\label{zeta*R}
\ee

\noindent{\it Non-resonant ICS}

\be 
\zeta _{\ast }^{(NR)} = 
\left\{ \begin{array}{ll}
    3~P^{7/6}/B_{12}^{2/3} = 0.3~B_{12}^{1/2}\tau _6^{7/12} & {\rm if}\: P\lsim P_{\ast }^{(NR)}, \\
    6~P^{7/4} /B_{12} = 0.2~B_{12}^{3/4}\tau _6^{7/8} & {\rm if}~P\gsim P_{\ast }^{(NR)},
\end{array} 
\right.
\label{zeta*NR}
\ee
where $\tau_6 = \tau/10^6~{\rm yr} = 65(P/B_{12})^2$.
Inserting expressions (\ref{zeta*CR}) - (\ref{zeta*NR}) into formulae 
(\ref{Lprim}), (\ref{fprim}) we get 

\noindent{\it Curvature radiation}

\be
L_{\rm prim}^{(CR)} = 10^{32}~{\rm erg/s}~~ 
\left\{ \begin{array}{ll}
    0.3~P^{-27/14}B_{12}^{6/7} = 20~B_{12}^{-15/14}\tau _6^{-27/28} & {\rm if}\: P\lsim P_{\ast }^{(CR)}, \\
    0.1~P^{-9/4}B_{12} = 10~B_{12}^{-5/4}\tau _6^{-9/8} & {\rm if}~P\gsim P_{\ast }^{(CR)},
\end{array} 
\right.
\ee
or
\be
L_{\rm prim}^{(CR)} = 10^{16}~{(\rm erg/s)^{1/2}}~~{\dot E}_{\rm rot}^{1/2}~~ 
\left\{ \begin{array}{ll}
    P^{1/14}B_{12}^{-1/7} & {\rm if}\: P\lsim P_{\ast }^{(CR)}, \\
    0.3~P^{-1/4} & {\rm if}~P\gsim P_{\ast }^{(CR)},
\end{array} 
\right.
\ee
\be
f_{\rm prim}^{(CR)} = 0.01  
\left\{ \begin{array}{ll}
    7~B_{12}^{13/14}\tau _6^{29/28} & {\rm if}\: P\lsim P_{\ast }^{(CR)}, \\
    5~B_{12}^{3/4}\tau _6^{7/8} & {\rm if}~P\gsim P_{\ast }^{(CR)},
\end{array} 
\right.
\ee

\noindent{\it Resonant ICS}

\be
L_{\rm prim}^{(R)} = 10^{31}~{\rm erg/s}~~ 
\left\{ \begin{array}{ll}
    0.3~P^{-13/6}=30~B_{12}^{-13/6}\tau _6^{-13/12} & {\rm if}\: P\lsim P_{\ast }^{(R)}, \\
    0.05~P^{-11/4}B_{12}^{1/2}=20~B_{12}^{-9/4}\tau _6^{-11/8} & {\rm if}~P\gsim P_{\ast }^{(R)},
\end{array} 
\right.
\ee
or
\be
L_{\rm prim}^{(R)} = 10^{15}~{(\rm erg/s)^{1/2}}~~{\dot E}_{\rm rot}^{1/2}~~ 
\left\{ \begin{array}{ll}
    P^{-1/6}B_{12}^{-1} & {\rm if}\: P\lsim P_{\ast }^{(R)}, \\
    0.2~P^{-3/4}B_{12}^{-1/2} & {\rm if}~P\gsim P_{\ast }^{(R)},
\end{array} 
\right.
\ee
\be
f_{\rm prim}^{(R)} = 10^{-2}  
\left\{ \begin{array}{ll}
    B_{12}^{-1/6}\tau _6^{11/12} & {\rm if}\: P\lsim P_{\ast }^{(R)}, \\
    0.8~B_{12}^{-1/4}\tau _6^{5/8} & {\rm if}~P\gsim P_{\ast }^{(R)},
\end{array} 
\right.
\ee

\noindent{\it Non-resonant ICS}

\be
L_{\rm prim}^{(NR)} = 10^{30}~{\rm erg/s}~~ 
\left\{ \begin{array}{ll}
    0.3~B_{12}^{2/3}P^{-13/6}=30~B_{12}^{-3/2}\tau _6^{-13/12} & {\rm if}\: P\lsim P_{\ast }^{(NR)}, \\
    0.1~B_{12}P^{-11/4}=30~B_{12}^{-7/4}\tau _6^{-11/8} & {\rm if}~P\gsim P_{\ast }^{(NR)},
\end{array} 
\right.
\ee
or
\be
L_{\rm prim}^{(NR)} = 10^{15}~{(\rm erg/s)^{1/2}}~~{\dot E}_{\rm rot}^{1/2}~~ 
\left\{ \begin{array}{ll}
    0.1~B_{12}^{-1/3}P^{-1/6} & {\rm if}\: P\lsim P_{\ast }^{(NR)}, \\
    0.05~P^{-3/4} & {\rm if}~P\gsim P_{\ast }^{(NR)},
\end{array} 
\right.
\ee
\be
f_{\rm prim}^{(NR)} = 10^{-3}  
\left\{ \begin{array}{ll}
    B_{12}^{1/2}\tau _6^{11/12} & {\rm if}\: P\lsim P_{\ast }^{(NR)}, \\
    2~B_{12}^{1/4}\tau _6^{5/8} & {\rm if}~P\gsim P_{\ast }^{(NR)}.
\end{array} 
\right.
\ee
The above formulae are just different representations 
of the luminosity that a primary (electron) beam achieves by the moment 
the $E_{\parallel}$ is screened, for each of the radiation processes 
we discuss. Note that the eight known EGRET $\gamma$-ray pulsars fit an 
empirical 
relation $L_{\gamma} \propto {\dot E}_{\rm rot}^{1/2}$.  This is reproduced by
the above expression for the luminosity of the primary beam at the
CR PFF if the primary electrons are nearly 100\% efficient at conversion
of their energy to high-energy $\gamma$-rays (see Zhang \& Harding 2000),
but not by the luminosity at the ICS pair fronts.
But we have argued that the primary particle acceleration is not
stopped at the ICS PFF, so that the potential drop at the CR PFF determines
the final acceleration energy and thus the $\gamma$-ray luminosity of
the pulsar.

\subsection{Derivation of theoretical pulsar death line}

\subsubsection{Analytic approach}

In pulsar physics the term ``death line" was originally 
introduced for radio pulsars and refers to a line separating 
the domain (normally in the $P$--$\dot P$ diagram) favouring 
pair formation from the domain where pair formation 
does not occur. Most PC models for radio 
pulsars imply therefore that radio emission turns off (see 
Sturrock 1971; Ruderman \& Sutherland 1975; and Chen \& 
Ruderman 1993) if the potential drop required for pair 
production exceeds the maximum which can be achieved over 
the PC. This concept of death line implicitly assumes 
that the pair-formation is a necessary condition for 
pulsar radio emission, and that pulsars become radio quiet 
after crossing the death lines during their evolution. 
Some authors argue that this condition may not be sufficient, 
though. For example, HA01 define a pulsar's 
death as the condition when pair-production is too weak to 
generate the pair cascade multiplicity required to screen the accelerating 
electric field. We believe that pair-formation is vital for pulsar 
operation, but we suggest that the pairs required for radio emission
may not 
necessarily screen or shut off the accelerating electric field (see also Paper I). 
In addition, we adopt a canonical definition of a pulsar's death 
as an essentially pairless regime of operation. In general, to 
write a formal criterion specifying the death line one needs 
to know the characteristic voltage drop in the pulsar acceleration region. 
For the acceleration model we employ in our study this 
voltage drop can be calculated as 
\be
\Delta \Phi _{acc} = 10^{10}~~{\rm V}~~B_{12} P^{-5/2} \zeta 
    \left\{ \begin{array}{ll}
    3~\zeta & {\rm if}\: P\lsim P_{\ast }^{(CR,R,NR)}, \\
    2 & {\rm if}~P\gsim P_{\ast }^{(CR,R,NR)}. 
\end{array} 
\right.
\label{phi_acc}
\ee
The corresponding calculated Lorentz-factor of a primary 
electron accelerating from the PC is 
\be
\gamma _{acc}  =  {e \over {mc^2}} \Delta \Phi _{acc} = 
10^4 B_{12} P^{-5/2} \zeta  
    \left\{ \begin{array}{ll}
    6~\zeta & {\rm if}\: P\lsim P_{\ast }^{(CR,R,NR)}, \\
    4 & {\rm if}~P\gsim P_{\ast }^{(CR,R,NR)}.
\end{array} 
\right.
\label{g_acc}
\ee

Thus, our criterion for the pulsar death line separating 
radio-active from radio-quiet pulsars translates into   
\be
\gamma _{acc} \geq \gamma _{\rm min} .      
\label{deathline}
\ee
However, the above expression for $\gamma _{acc}$, 
for fixed values of $B$ and $P$, still depends 
on a dimensionless acceleration altitude, $\zeta $, 
and can not be used therefore in criterion (\ref{deathline}).
It is important to note, that in our numerical calculations, 
at each altitude, we check the instant value of a particle's 
Lorentz-factor against the above criterion. If it meets this 
criterion, we pick the corresponding bulk pulsar 
parameters such as $B$ and $P$ and store them 
as those of the death line. Generally, 
to make our analytic derivation of the death line  
self-consistent, we need an additional independent equation 
that relates $\zeta $ with e.g. $B$ and $P$. Obviously, the 
formal substitution of values of $\zeta _{\ast}$ calculated in 
\S~2.2 into $\gamma _{acc}$ in criterion (\ref{deathline}) 
would lead to a trivial identity, simply because in our 
derivation of $\zeta _{\ast }$ we 
have already used the corresponding values of $\gamma _{\rm min}$. 
Instead, we suggest using equations (\ref{fprim}) and (\ref{g_acc})
to eliminate $\zeta $ to get 
\be   
\gamma _{acc} \approx 6\cdot 10^{7} f_{\rm prim} B_{12} P^{-2} .
\label{gamma_acc}
\ee
The advantage of this formula for $\gamma _{acc}$ is that it does 
not discriminate between unsaturated and saturated regimes of 
acceleration of primary electrons and is factorized by the efficiency 
of the pulsar accelerator, $f_{\rm prim}$, which may be generally regarded 
as a free pulsar parameter, characterizing the efficiency of primary 
acceleration above the pulsar PC. 

Now, let us use formula (\ref{gamma_acc}) in criterion 
(\ref{deathline}) to get explicit expressions describing the death 
lines for different underlying radiation mechanisms for pair-producing 
photons. Note that for this purpose in criterion (\ref{deathline}) we 
should evaluate $\gamma_{acc}$ at $f_{\rm prim}=f_{\rm prim}^{\rm min}$, where 
$f_{\rm prim}^{\rm min}$ is the minimum pulsar efficiency allowing pair formation.  
This completes our formal definition of death lines and makes it physically 
sensible: for the fixed values of $B$ and $P$, to specify the onset of pair 
formation, one needs to know the minimum energetics a primary electron beam 
should have to enable pair formation, i.e. one needs to know $f_{\rm prim}^{\rm min}$. 
We must note that, within the framework of this approach, the analytic death 
lines derived in previous studies implicitly assume some fixed value for 
$f_{\rm prim}^{\rm min}$, the same for all pulsars and for all relevant 
pair-formation mechanisms, and that those values of $f_{\rm prim}^{\rm min}$ are not 
necessarily meaningful. On the contrary, our analytic death lines are explicitly 
determined by a pulsar bulk parameter $f_{\rm prim}^{\rm min}$. The resultant analytic 
death lines (or rather parameter spaces with allowed pair formation) in 
the $P$--$\dot P$ diagram read 

\noindent{\it Curvature radiation}

\be 
\log {\dot P} \geq 
\left\{ \begin{array}{ll}
    {21\over 8}~\log P - 1.75\log f_{\rm prim}^{\rm min} - 16 & {\rm if}\: P\lsim P_{\ast }^{(CR)}, \\
    {5\over 2}~\log P - 2\log f_{\rm prim}^{\rm min} - 17 & {\rm if}~P\gsim P_{\ast }^{(CR)},
\end{array} 
\right.
\label{DL_CR}
\ee

\noindent{\it Resonant ICS}

\be 
\log {\dot P} \geq 
\left\{ \begin{array}{ll}
    {5\over 6}~\log P - \log f_{\rm prim}^{\rm min} - 17.4 & {\rm if}\: P\lsim P_{\ast }^{(R)}, \\
    {2\over 3}~\log P - 1.33\log f_{\rm prim}^{\rm min} - 18.9 & {\rm if}~P\gsim P_{\ast }^{(R)},
\end{array} 
\right.
\label{DL_R}
\ee

\noindent{\it Non-resonant ICS}

\be 
\log {\dot P} \geq 
\left\{ \begin{array}{ll}
    {7\over 4}~\log P - 1.5\log f_{\rm prim}^{\rm min} - 19.8 & {\rm if}\: P\lsim P_{\ast }^{(NR)}, \\
    {3\over 2}~\log P - 2\log f_{\rm prim}^{\rm min} - 21.6 & {\rm if}~P\gsim P_{\ast }^{(NR)}.
\end{array} 
\right.
\label{DL_NR}
\ee
The fitting of observational data with the above expressions requires specification 
of the pair-producing efficiency ($f_{\rm prim}^{\rm min}$) of a pulsar PC accelerator. In 
this paper we calculate pulsar death lines numerically, and by comparing the 
analytic death lines with our numerical death lines we can estimate the values 
of $f_{\rm prim}^{\rm min}$ (see next Section). It is of fundamental importance to know 
if this value of $f_{\rm prim}^{\rm min}$ (i.e. $f_{\rm prim}$ calculated along a pair  
death line) is the same for all pulsars, and how it depends on the radiation 
process responsible for the pair formation. We find that the analytic death lines 
defined by (\ref{DL_CR})-(\ref{DL_NR}) with the appropriately chosen value of 
$f_{\rm prim}^{\rm min}$, different for each radiation mechanism, fit satisfactory the 
observational data and are in a good agreement with our numerically 
calculated death lines. This means that our introduction of the parameter 
$f_{\rm prim}^{\rm min}$ and the assumption that this parameter only weakly depends on 
pulsar $B$ and $P$ values, for a given mechanism of generation of pair-producing 
photons, may be quite justified for the analytic tackling of pulsar death lines.

The analytic expressions above for the pair death lines differ significantly from
those derived by Zhang et al. (2000).  The pair death lines of Zhang et al. (2000)
were derived from the condition (in our notation) that $\Delta\Phi_{acc} = \Delta\Phi_{max}$,
which implicitly assumes that $f_{prim}^{min} = 1$, whereas we allow $f_{prim}^{min}$
to be a free parameter.  As we shall see in the next section, comparison with numerical
calculations implies $f_{prim}^{min} \ll 1$.  Also, Zhang et al. (2000) computed the
altitude of the PFF and thus the value of $\Delta\Phi_{acc}$ from the
condition $S_p (\varepsilon _{\rm min}) = S_e(\gamma_{\rm min})$, that photon attenuation length
is equal to the length required for the primary electron to produce one ICS photon,
since the acceleration length $S_{acc} \ll S_p$.
This introduced a dependence of the PFF altitude on PC temperature, which they took
to be the self-consistent temperature from PC heating.  By contrast, our analytic
expression for the PFF altitude (eq. [\ref{S0}]) does not depend on PC temperature.
Finally, Zhang et al. (2000) included pairs from only RICS and did not include pairs
from NRICS in their ICS death lines.

\subsubsection{Comparison with numerical calculations}

The discussion and results of our numerical calculation of the 
CR and ICS pair death lines were presented in Figure 1.  Since
our numerical solution for the location of the PFFs are computed by
iteration, we can unambiguously determine the pair death lines
without having to define the efficiency,
$f_{\rm prim}^{\rm min}$, as was needed in the analytic expressions above.  

We must note that in our derivation of the above analytic death-line 
conditions we did not require the value of accelerating 
potential drop to be maximum at the PFF. In other words, we did 
not impose any {\it ad~hoc} constraint on the acceleration 
altitude, thus making the latter consistent with the 
accelerating potential drop. It is remarkable that for each of 
the radiation mechanisms the parameter $f_{\rm prim}^{\rm min}$ 
(minimum efficiency of acceleration of primary beam required to 
set pair formation) remains nearly constant along 
the numerical death line.  By comparing our numerically 
computed death lines with those given by 
(\ref{DL_CR})-(\ref{DL_NR}) we find that $f_{\rm prim}^{\rm min}  
\sim 0.03-0.1$ for CR, 0.003 - 0.02 for RICS and $4 - 6 \times 10^{-3}$
for NRICS pair fronts. These efficiencies are still small compared 
to the maximum efficiency we esimate in the last paragraph of this 
Section.  Another important finding is that production of pairs in
all observed pulsars by CR photons would require 
$f_{\rm prim}^{\rm min} \gg 1$ indicating the difficulty in interpreting 
the observational data in terms of a CR mechanism alone. On the contrary,  
ICS-based death lines imply much less consumption of 
pulsar spin-down power for pair creation. This result and 
our finding that the ICS-generated pairs tend to only 
partially screen the accelerating electric field may 
suggest the occurrence of ``nested'' pair formation regions: the 
lower-altitude pairs produced by the ICS photons and the 
higher-altitude pairs produced by the CR photons generated 
by particles accelerating through the region of the ICS pair 
formation. 

This fact implies that if a pulsar is below the CR death line in Figure 1, 
and if any ICS screening is inefficient, then the primary beam acceleration 
is not limited by pair production and will be producing ample 
CR high-energy photons. The upper limit for the pulsar $\gamma $-ray 
luminosity in this case can be estimated by using equation (\ref{Lprim_gen}) 
with the integrand evaluated at the maximum value of the potential given 
by formula (13) of HM98 (see also eq. [A4] of HM98):
\be
L_{\gamma , {\rm max}} = c \int _{S(\eta )}\alpha (\eta )\rho (\eta, \xi) 
\Phi _{\rm max}dS(\eta ), 
\label{Lgamma}
\ee
where $\Phi _{\rm max} \approx (1/2)[\Omega R/cf(1)]\kappa \Phi _0$, 
$\Phi_0 = (\Omega R/c)B_0 R$, and $dS(\eta ) = S(\eta) 
d\Omega _{\xi }/\pi $, where $S(\eta )$ is the spherical cross-sectional 
area of the polar flux tube at radial distance r ($=\eta R$). Note that in 
formula (\ref{Lgamma}) the integral $c\int \alpha \rho dS $
represents the total electron current flowing from the PC and is a constant.   
Thus, integrating over $\xi $ and $\phi $, we arrive at 
\be
L_{\gamma , {\rm max}} \approx 1.5 \pi \kappa (1-\kappa ) 
{\dot E}_{\rm rot} \approx 0.6 {\dot E}_{\rm rot}.
\label{Lgamma_max}
\ee
Here we used $\kappa = 0.15$ and $\cos \chi \approx 1$. 
We predict, therefore, that pulsars tend to be efficient $\gamma $-ray 
sources if they are to the right of their CR death lines in B-P diagram. 
Note that expression (\ref{Lgamma_max}) can also be used to set the possible 
upper limit for $\kappa $, which is around 0.3.


\subsection{Characteristic voltage drop at the pair front}

Now we would like to demonstrate one more remarkable property 
of a pulsar PC accelerator.  Let us take the expression for the 
electric potential as a function of $\zeta $ 
(see eq. [\ref{phi_acc}]) and evaluate it at the screening altitude
(at $\zeta = \zeta _{\ast }$). After substituting the 
corresponding expressions for $\zeta _{\ast }$ 
(see eqs. [\ref{zeta*CR}]-[\ref{zeta*NR}]) into equation 
(\ref{phi_acc}) we get the following self-consistent 
formulae for the critical voltage drop (characteristic 
voltage drop at the screening altitude)
 
\noindent{\it Curvature radiation}

\be 
\Delta \Phi _{acc}(\zeta _{\ast }) = 10 ^{13}~{\rm V}~~ 
\left\{ \begin{array}{ll}
    2~(\tau_6/P)^{1/14} & {\rm if}\: P\lsim P_{\ast }^{(CR)}, \\
    P^{-1/4} & {\rm if}~P\gsim P_{\ast }^{(CR)}.
\end{array} 
\right.
\ee

\noindent{\it Resonant ICS}

\be 
\Delta \Phi _{acc}(\zeta _{\ast }) = 10 ^{11}~{\rm V}~~ 
\left\{ \begin{array}{ll}
    3~P^{-7/6}\tau_6^{1/2} & {\rm if}\: P\lsim P_{\ast }^{(R)}, \\
    2~P^{-5/4}\tau_6^{1/4} & {\rm if}~P\gsim P_{\ast }^{(R)},
\end{array} 
\right.
\ee

\noindent{\it Non-resonant ICS}

\be 
\Delta \Phi _{acc}(\zeta _{\ast }) = 10 ^{11}~{\rm V}~~ 
\left\{ \begin{array}{ll}
    2~P^{-1/2}\tau_6^{1/6} & {\rm if}\: P\lsim P_{\ast }^{(NR)}, \\
    P^{-3/4} & {\rm if}~P\gsim P_{\ast }^{(NR)}.
\end{array} 
\right.
\ee
These formulae show very weak dependence of critical voltage drop 
on pulsar parameters $B$ and $P$, especially for the CR case, as
also shown by HM98. 
This fact simply indicates that the critical voltage drop 
needed for the ignition of a pair formation is mainly  
determined by the pair-formation microphysics itself. As a result, 
as we move from the lower left to the upper right corner of 
the $P$--$\dot P$ diagram, the variation (by orders of magnitude) 
in the altitude of the PFF effectively compensates the corresponding 
change in  voltage drop due to its scaling with pulsar parameters $B$ 
and $P$, thus maintaining it near its critical value. 

Note that in similar expressions derived by HA01 (see 
their eqs. [73]-[75]) the characteristic voltage drop varies 
by $\sim $ four orders of magnitude over the whole range of 
pulsar spin periods $\sim 0.001-10$ s. 

\section{Polar cap heating luminosities and surface temperatures}

\subsection{Analytic estimates}

To estimate the efficiency of PC heating 
by returning positrons, we can use formulae 
(64), (65) of Paper I again evaluated at 
$z_{\ast }$. 

Thus, setting $\kappa = 0.15$ and $\cos \chi \approx 1$, 
we get for the heating efficiency, $f_+ = L_+ / \dot E_{rot}$,

\noindent{\it Resonant ICS}

\be  \label{f+R}
f_+^{(R)} = 10 ^{-5} P^{-1/2}\tau _6^{3/2}  
\left\{ \begin{array}{ll}
    2.5 & {\rm if}\: P\lsim P_{\ast }^{(R)}, \\
    1.5 & {\rm if}~P\gsim P_{\ast }^{(R)},
\end{array} 
\right.
\ee

\noindent{\it Non-resonant ICS}

\be \label{f+NR}
f_+^{(NR)} = 10 ^{-5} P^{1/2}\tau _6  
\left\{ \begin{array}{ll}
    1.0 & {\rm if}\: P\lsim P_{\ast }^{(NR)}, \\
    0.6 & {\rm if}~P\gsim P_{\ast }^{(NR)}.
\end{array} 
\right.
\ee
Here we shall summarize the explicit expressions 
for the estimated PC luminosities and surface temperatures due 
to the heating by positrons returning from the pair fronts set 
by the CR and ICS photons. Our estimates of the PC temperature  
are based on a standard formula 
$T_{pc} \approx (L_+/\pi r_{pc}^2\sigma _{SB} )^{1/4}$, 
where $\sigma _{SB} $ is the Stefan-Boltzmann constant.  
This formula implies therefore that the PC is heated 
homogeneously, and that the heated area is confined by the 
PC radius. 

\noindent{\it Curvature radiation}

\be 
L_+^{(CR)} = 10 ^{31}~{\rm erg/s}~P^{-1/2}  
\left\{ \begin{array}{ll}
    0.4~P^{-5/14}\tau _6^{-1/7} & {\rm if}\: P\lsim P_{\ast }^{(CR)}, \\
    1.0 & {\rm if}~P\gsim P_{\ast }^{(CR)},
\end{array} 
\right.
\label{L+CR}
\ee
\be 
T_{pc}^{(CR)} = 10 ^6~{\rm K}~P^{-1/4}  
\left\{ \begin{array}{ll}
    P^{-5/56}\tau _6^{-1/28} & {\rm if}\: P\lsim P_{\ast }^{(CR)}, \\
    2 & {\rm if}~P\gsim P_{\ast }^{(CR)},
\end{array} 
\right.
\label{TCR}
\ee

\noindent{\it Resonant ICS}

\be 
L_+^{(R)} = 10 ^{28}~{\rm erg/s}~P^{-5/2} \tau _6^{1/2}   
\left\{ \begin{array}{ll}
    0.8 & {\rm if}\: P\lsim P_{\ast }^{(R)}, \\
    0.5 & {\rm if}~P\gsim P_{\ast }^{(R)},
\end{array} 
\right.
\label{L+R}
\ee
\be 
T_{pc}^{(R)} \approx 8\times 10 ^3~{\rm K}~P^{-9/4} \tau _6^{1/8} ,
\label{TR}
\ee

\noindent{\it Non-resonant ICS}

\be \label{L+NR}
L_+^{(NR)} = 10 ^{28}~{\rm erg/s}~P^{-3/2}
\left\{ \begin{array}{ll}
    0.3 & {\rm if}\: P\lsim P_{\ast }^{(NR)}, \\
    0.2 & {\rm if}~P\gsim P_{\ast }^{(NR)},
\end{array} 
\right.
\ee
\be 
T_{pc}^{(NR)} \approx 4\times 10 ^4~{\rm K}~P^{-5/4} .
\label{TNR}
\ee
We caution that the above formulae are upper limits to the PC heating
luminosity, in that they assume locally complete screening.  As we have shown 
in Section 3, such screening does not occur in many cases, most notably
for older pulsars.  Thus as will be shown by the numerical calculation presented
in the next Section, eqs. (\ref{L+NR}) and (\ref{TNR}) do not
accurately predict $L_+$ and $T_{pc}$ for ms pulsars, but give values that are 
much too high.

Let us compare the above temperature estimates with those of HA01. 
At $B_{12} = 1$ and $P = 0.1$ s our estimate of $T_{pc}^{CR}$ is
roughly the same as the corresponding estimate presented by HA01 (see 
their eqs. [58]). For the ICS cases (see their eqs. [56], [57] and 
[59], [60]) their estimates are systematically bigger, by a factor of 3 
to 12. However, for the ms pulsars we predict higher temperatures.  Also,  
the effect that the ICS contribution to the PC temperatures dominates over 
the CR contribution for ms pulsars is more pronounced in our formulae. In 
addition, for long-period pulsars our formulae predict more drastic decline 
in the PC heating temperatures with a pulsar period.  We also comment that
both the analytic estimates and numerical calculations of $L_+$ depend
on the value of $\kappa$, related to the compactness and moment of 
inertia of the NS.

\subsection{Numerical calculations}

We numerically evaluate the luminosity of PC heating due to ICS PFFs
in a similar way to that given by equation (62) of Paper I.  The fraction of returning 
positrons times the potential drop between the PFF and the surface is integrated
in $\xi$ across the PC.  However, since in the case of ICS PFFs the 
potential varies significantly across the width of the PFF, we also integrate 
the product of the potential and the returning positron fraction at each 
altitude between $z_0$ and $z_0 + \Delta_s$:
\be
L_+ = 2 \alpha c S_{pc} \eta _0^3 {{f(1)}\over {f(\eta _0)}} \int _0^1 
\int_{z_0}^{z_0 + \Delta_s}  \rho _+(z-z_0,\xi) 
\Phi (z,\xi )\,dz\,\,\xi d\xi = \int _0^1 L_+(\xi)\,\xi d\xi 
\label{L_+II}
\ee
where $S_{pc} = \pi \Omega R^3 /c f(1)$ is the area of the PC.  Figure 7 shows 
the dependence of $L_+ (\xi)$ on $\xi$, which reflects the distribution of heating 
across the PC.   Most of the PC heating due to the ICS pair front occurs
at small $\xi$, near the magnetic pole for normal pulsars, but in the outer
part of the PC for ms pulsars.

The total positron heating luminosity, scaled with the spin-down luminosity, as a 
function of characteristic pulsar age, $\tau = P/2\dot P$ is shown in Figure 8 for two
different values of PC temperature.  The numerically computed $L_+/\dot E_{\rm rot}$ 
increases with $\tau$, in agreement with the analytic estimate (eqs. [\ref{f+R}]
and [\ref{f+NR}]), as long 
as there is locally complete screening.  When complete screening is no longer
achieved, at a $\tau$ value that depends on $P$ and $T_6$, the heating
rate saturates because the number of pairs produced is not sufficient to return a
positron flux proportional to $z_0$.  Even at very high PC temperatures, the ms pulsars
have reached saturation of the PC heating rate because they are beyond the ICS 
screening boundary.  The heating luminosity is larger for a lower value of $T_6$
before saturation because the PFF is at higher altitude and the potential drop
is higher, giving the positrons more energy before they reach the NS surface.  
Figure 9 displays the same calculations of the PC heating
luminosity, not scaled to the spin-down luminosity, to more easily compare with
observations.  In these Figures we have also plotted for comparison the calculations 
of PC heating luminosity from the CR PFFs presented in Paper I.  
The high $\tau$ end of the CR 
heating rate curves marks the CR PFF pair death line for that period.  Where heating from
positrons returning from the CR PFF is present, it is several orders of magnitude
larger than the heating from the ICS PFF, so that in these pulsars heating from the ICS
pairs makes a negligible contribution to the total PC heating rate.  For normal pulsars,
heating from ICS pair fronts is not high enough to be detectable at present,
even in the absence of CR pair heating.  However, emission from ICS pairs provides
the only source of PC heating for the known ms pulsars, which have ages 
$\tau > 10^8$ yr,
since they cannot produce CR pairs.  This emission may be detectable if the PC 
temperatures are above $10^6$ K.  The relatively high ICS pair heating rates in
the ms pulsars is due primarily to the higher voltage drop necessary to make pairs.
Even though the fraction of returning positrons is not large, the positrons gain more
energy before hitting the NS surface.

The analytic estimates for $L_+$ are in reasonably good
agreement with our numerical calculations (within a factor of 10) in the 
regime where locally
complete screening is occurring.  Where compete screening is not achieved, the
analytic estimates of $L_+$ and also of $T_{pc}$ given in Section 5.1 will greatly 
overestimate the true values.  This is especially true in the case of the ms
pulsars.  We caution that equations (\ref{L+R}) - (\ref{TNR})
cannot be used for pulsars that are beyond the screening boundaries of Figure 4.  

In Figure 10, we show the predicted luminosities from ICS PC heating, $L_+$ as a function of 
PC surface temperature (solid lines) for two different pulsar periods and a surface field of
$B_0 = 4.4 \times 10^8$ G, so that the corresponding characteristic ages would be
$\tau = 1.4 \times 10^9$ yr for P = 2 ms and $\tau = 8.75 \times 10^9$ yr for P = 5 ms.
There is a definite dependence $L_+ \propto T^2$ seen in these curves, which cannot
be modeled analytically since these cases do not have full screening.
Also plotted (dashed lines) are the luminosities, $L_{BB} = A\sigma_{SB} T^4$, emitted by 
a blackbody at PC radiating temperature 
$T$ and different heated surface areas $A$.  The intersections of the
curves roughly indicate values of temperature where self-sustained heating,
(i.e. where the surface emission at a given temperature supplies just enough returning
positrons from ICS pairs to maintain the PC at that temperature)
is possible for heated PCs of a particular area.  Since the standard PC area
is $A_{pc} = 3.3 \times 10^{11}\,{\rm cm^2}$ for P = 2 ms and $A_{pc} = 1.3 \times 
10^{11}\,{\rm cm^2}$ for P = 5 ms, self-sustained PC heating emission requires
heated and radiating areas much smaller than the entire PC area.  The intersection
points of the blackbody and $L_+$ curves in Figure 10 are not entirely self-consistent
self-sustained heating models since we assume a radiating area of $A_{pc}$ for 
our calculations.  The fully self-consistent models would have radiating areas 
somewhat larger at a given temperature, but still smaller than $A_{pc}$.

Of the six millisecond radio pulsars which have been detected as pulsed X-ray sources,
most have narrow pulses and power-law spectra indicating that their emission is 
dominated by non-thermal radiation processes.  However, two pulsars,
PSR J0437-4715 (P = 5.75 ms, $\tau = 4.6 \times 10^9$ yr) and PSR J2124-3358 
(P = 4.93 ms, $\tau = 7.3 \times 10^9$ yr), have possible thermal emission components 
which would imply that some surface heating is taking place.  The emission from 
PSR J0437-4715 has a dominant power-law component but a two-component model is
required for an acceptable fit.  A two-component power-law plus blackbody fit to combined
EUVE and ROSAT data (Halpern et al. 1996) give a temperature of $T = (1.0 - 3.3)
\times 10^6$ K, luminosity $L = 8.4 \times 10^{29}\,{\rm erg\,s^{-1}}$ and
emitting area $A = 7.8 \times 10^7 - 1.1 \times 10^{10}\,{\rm cm^2}$ for the
thermal component.  Recent Chandra observations of PSR J0437-4715 (Zavlin et al. 2001) 
confirm that at least one thermal component plus a power law is needed to fit 
the spectrum.  Their preferred model consists of a two-temperature thermal blackbody,
with a hotter PC core of $T_{core} = 2.1 \times 10^6$ K and $R_{core} = 0.12$ km and
a cooler PC rim of $T_{rim} = 5.4 \times 10^5$ K and $R_{rim} = 2.0$ km.  
A blackbody fit to ASCA emission from PSR J2124-3358 (Sakurai
et al. 2001) yields a temperature $3.6^{+0.93}_{-0.70} \times 10^6$ K, luminosity 
$L = 4.8 \times 10^{29}\,{\rm erg\,s^{-1}}$ and PC emitting area $A = 1.4^{+2.5}_{-0.9}
\times 10^7\,{\rm cm^2}$.  

Comparing our results in Figure 10 with the observed values of $T$, $L$ and $A$ for
PSR J0437-4715 and PSR J2124-3358, we see that the predicted luminosities,
areas and temperatures for self-sustained PC heating are in a range comparable to 
observed values.  Thus ICS pair fronts could be a plausible source of PC heating
for some ms pulsars.

Recent Chandra high-resolution X-ray observations of the galactic globular cluster 47~Tuc
(Grindlay et al. 2001a) have detected all 15 known ms radio pulsars in the cluster
as well as a number of other suspected ms pulsars.  It is thought that the X-ray
source population of 47 Tuc is dominated by ms pulsars having soft spectra and 
luminosities $L_X \sim 10^{30}\,{\rm erg\,s^{-1}}$.  
Grindlay et al. (2001b) have found a dependence
$\log(L_X) = -(0.32\pm 0.1)\log(\tau ) + 33.3$ for the ms pulsars in the cluster,
not too different from what we have found for PC heating from ICS pair fronts (cf.
Figure 9b).  

\section{SUMMARY AND CONCLUSIONS}

We have explored production of electron-positron pairs by photons produced through
ICS of thermal X-rays by accelerated electrons above a pulsar 
PC.  Since the accelerating primary electrons can produce pairs from ICS photons
at much lower energies than are required to produce pairs from CR photons, it is
very important to investigate the consequences of ICS pair fronts for $E_{\parallel}$
screening and for PC heating.  We have defined ``pair death" lines in 
$B_0 - P$ space as the boundary of pair production for pulsars, having dipole magnetic fields
of surface strength $6.4 \times 10^{19}(\dot P P)^{1/2}$ G.  
Operationally, the existence of a pair front 
is determined by a finite solution to equation (\ref{S0}) for $S_0$, the altitude of 
the onset of pair creation.  Although we are able to give analytic formulae for the 
altitude of the PFFs for different radiation processes, the 
location of the pair death lines must be determined numerically. 
The existence of a pair front requires much less than one pair per primary electron
(but still many pairs from the whole PC beam),
since the very first pairs are created in the declining high-energy tail of the radiated 
spectrum, and also much less than the number of pairs per primary required for screening
of $E_{\parallel}$.  We find that virtually all known radio pulsars are capable of
producing pair fronts with ICS photons.  A smaller number, less than half, are able
to also produce pairs via CR.  If the acceleration model we use is correct and pulsars
have dipole fields, then this result implies that relatively few pairs are required
for coherent radio emission.  

Self-consistent calculations show that ICS pair fronts produce lower fluxes of returning 
positrons and lower PC heating luminosities than CR pair fronts.  This is due to
the higher efficiency of the ICS process in producing pairs at lower altitudes where
both the charge deficit required to screen (and thus the returning positron flux) 
and the accelerating voltage drop are smaller.
For pulsars with surface magnetic fields in the ``normal" range of $10^{11} - 10^{13}$ G,
ICS heating luminosities are several orders of magnitude lower than CR heating luminosities.
However, for ms pulsars having surface fields in the range $10^8 - 10^{10}$ G, production
of any pairs requires such high photon energies that ICS pair fronts occur at higher 
altitudes, where acceleration voltage drops are high enough to produce significantly more
PC heating.  Since most ms pulsars cannot produce pairs through CR, ICS pair fronts 
provide the only means of external PC heating.  We find that for surface temperatures
$T \gsim 10^6$ K, ICS heating luminosities are in the range of detection.

We find that ICS pairs are able to screen the {\it local} $E_{\parallel}$ in some pulsars
having a high enough PC temperature, but that this local screening will not
produce a complete screening of the accelerating field at all altitudes and thus will
not stop the acceleration of the primary beam.  This is because the number of ICS pairs 
grows slowly,
on a scale length comparable to the altitude of the PFF, and then declines while the charge deficit which maintains $E_{\parallel}$ continues to increase with altitude.  Even if the
ICS pair production is vigorous enough to achieve local screening of $E_{\parallel}$,
it eventually cannot produce the charge density (from returning positrons) to keep up
with the increasing charge deficit (produced by the combination of flaring field lines
and inertial frame-dragging).  The primary particles may slow their acceleration briefly,
due to the local screening, but will resume acceleration once the ICS pair production 
declines.  At higher altitudes, those pulsars to the left of the CR pair death line 
will reach the Lorentz factors ($\gamma \sim 2 \times 10^7$) required to produce CR pairs.
In contrast to ICS pair fronts, the growth of pairs above the CR PFF is rapid and robust
due to the sensitivity of CR photon energy and emission rate on particle Lorentz factor, 
producing complete screening of $E_{\parallel}$ in a very short distance (cf. Paper I).  In 
pulsars to
the right of the CR pair death line, there is not complete screening and the primary particles 
will continue accelerating to high altitude with their Lorentz factor being possibly
limited by CR reaction.

We have also investigated the proposal by HM98 that pairs produced as the returning
positrons are accelerated toward the NS may be able to screen the $E_{\parallel}$
above the surface.  Using our calculated values of the returning positron flux
in cases where local screening has been achieved at the ICS pair fronts, we find that
in most cases screening does not occur at a significant distance above the NS surface 
to cause disruption of a steady state or formation of pair fronts.  
In the cases where screening does
occur near PFFs significantly above the surface, the pulsars are near the ICS screening 
boundary.  The resulting instability would then not move the start of the acceleration to
higher altitudes, as HM98 had envisioned, but would probably weaken or disrupt the
screening at the upper ICS pair front, resulting in a decrease in returning positron
flux which would weaken the screening at the lower PFF, etc.

PC heating by CR pair fronts will dominate for pulsars to the left of the CR pair
death line, while ICS pair fronts will supply the PC heating for pulsars to the right of
the CR pair death line.  While we have given analytic expressions for the fraction of
returning positrons and PC heating luminosities from ICS pair fronts, these are only good 
estimates above the ICS screening boundaries of Figure 4.  Below the ICS screening
boundaries, where local screening is not achieved for that PC temperature, the 
numerical values of returning positron fraction and heating luminosity fall well below  
the analytic estimates.  This will be true for ms pulsars, nearly all of which are below
the ICS screening boundary for a PC temperature of even $T = 10^7$ K. 

Our results are dependent on a number of assumptions inherent in our calculations.
First, we have assumed that the pattern of thermal X-ray emission is from a heated PC 
and is a pure, isotropic blackbody.  According to recent calculations (Zavlin et al.
1995) of radiation transfer in magnetized NS atmospheres, the thermal emission is not pure 
blackbody or isotropic,
but a somewhat cooler blackbody consisting of pencil and fan beam components.  Both 
effects of full surface (cooler) emission and pencil beaming would tend to decrease
ICS radiation and thus ICS pair production efficiency.  However, a large fan beam
component would tend to increase ICS efficiency.       
Second, we have used a hybrid scheme to describe the ICS radiation spectrum in which
the RICS has been treated as classical magnetic Thompson scattering and
NR (or continuum) ICS has been treated as relativistic but non-magnetic.
In reality, both RICS and NRICS should be treated as a single process with one 
cross section.  While the magnetic QED scattering cross
section has been studied for some time (e.g. Herold 1979, Daugherty \& Harding 1986), 
simple expressions in limited cases are only beginning to become available (e.g.
Gonthier et al. 2000).  Our present treatment is probably accurate for magnetic fields
$B \lsim 0.2 B_{\rm cr}$ which includes most of the radio pulsars. 

The location of a pulsar relative to the pair death lines may be important not only to
its radio and thermal X-ray emission characteristics, but also to its high-energy
emission properties.  As we have argued in this paper, ICS pair fronts will not limit
the acceleration voltage drop in pulsars but that acceleration will continue until it is
limited by a CR pair front.  The voltage drop at the CR pair front (together with the size of 
the PC current) is therefore expected to determine the high-energy emission luminosity.  
In Section 4, we have noted that the acceleration voltage drop of pulsars that produce CR
pair fronts is remarkably insensitive to pulsar parameters, leading to the prediction
that high-energy luminosity, $L_{\sc HE}$, should be simply proportional to PC current 
(which is proportional to $\dot E_{\rm rot}^{1/2}$), which
seems to be borne out by observations (e.g. Thompson 2000).  However, pulsars that do
not produce CR pair fronts do not have such a limit on acceleration voltage drop  and should
depart from the $L_{\sc HE} \propto \dot E_{\rm rot}{1/2}$ dependence, and approach
a $L_{\sc HE} \propto \dot E_{\rm rot}$ dependence.  Indeed, such a departure
must occur if they are not to exceed 100\% efficiency in converting rotational energy
loss to high-energy emission.  We predict that this change in $L_{\sc HE}$ dependence should
occur along the CR pair death line.  None of the pulsars which have detected $\gamma$-ray
emission are to the right of the CR pair death line although some, such as Geminga, are close.
The Large Area Gamma-Ray Telescope (GLAST) will have the sensitivity to detect $\gamma$-ray
emission from significant numbers of radio pulsars beyond the CR pair death line and so
should be able to test this prediction.

\acknowledgments 
We thank Josh Grindlay, Jon Arons, George Pavlov and Bing Zhang for comments and discussion. 
We also thank the referee for valuable comments.  This work was supported by  
the NASA Astrophysics Theory Program.

\newpage

\figureout{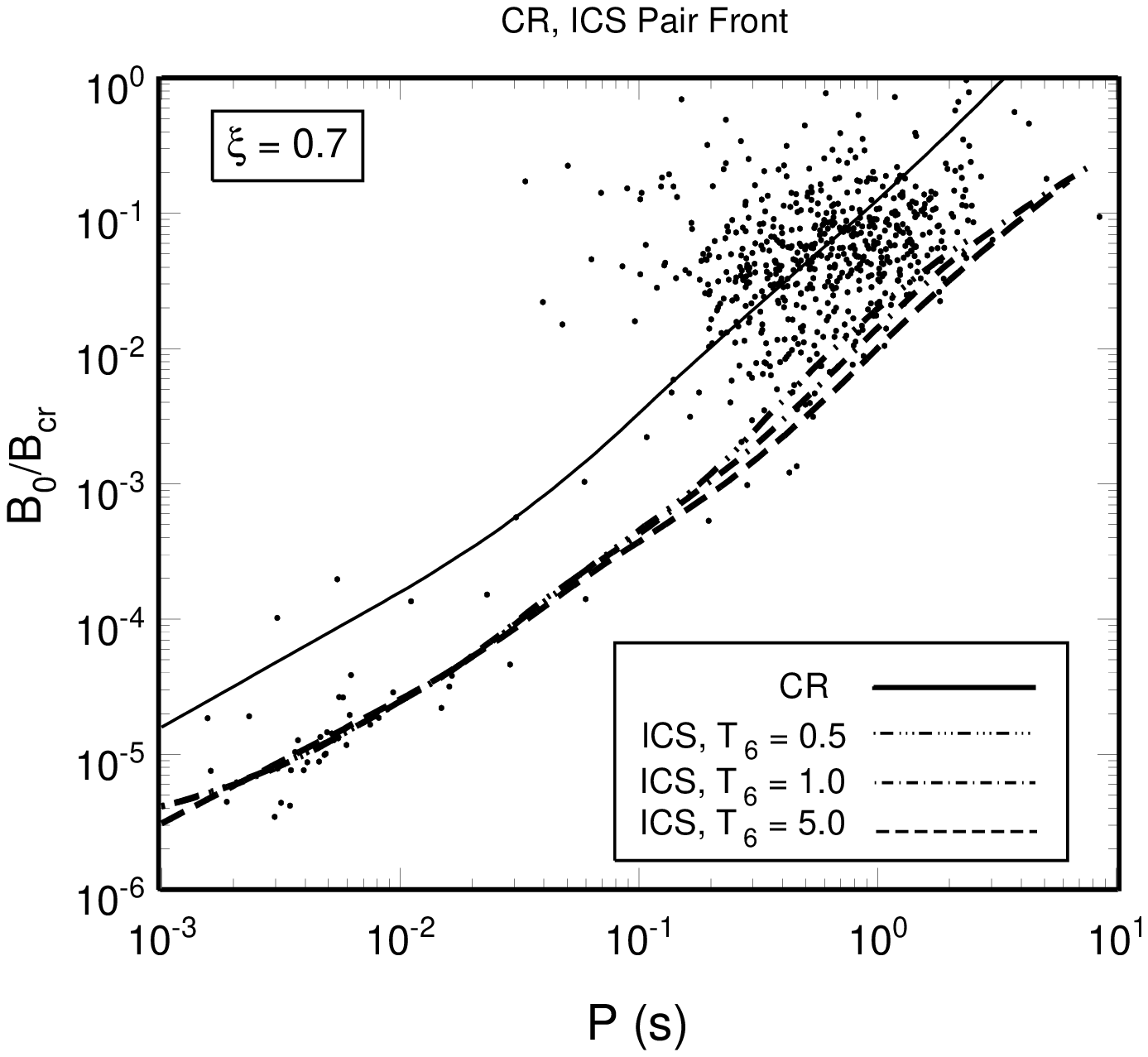}{0}{
Boundaries defining regions in surface magnetic field $B_0$, in units of critical field, 
versus pulsar period, $P$, where pulsars are capable of producing pairs through 
curvature radiation (CR) or inverse-Compton radiation (ICS).   The ICS curves are 
labeled with different values of PC temperature, $T_6$, in units
of $10^6$ K.  Also shown are radio pulsars from the ATNF Pulsar Catalog. 
Pulsars to the right of each line cannot produce 
pairs by the corresponding process.   
    \label{fig:PFF} }    

\figureout{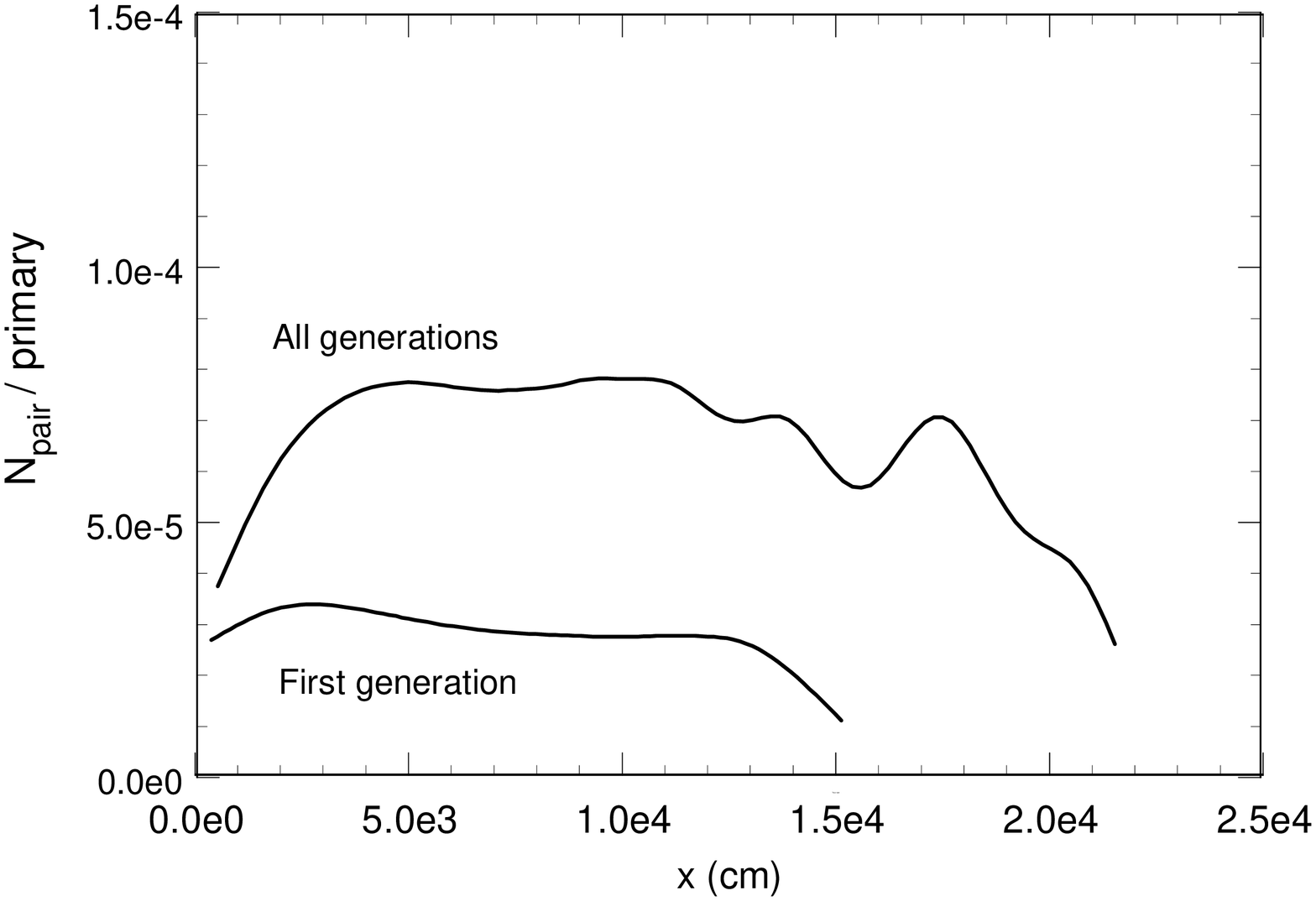}{0}{
Example of electron-positron pair source function integrated over energy, as a 
function of distance, x, above the pair formation front produced by a pair cascade 
initiated by inverse-Compton radiation from primary electrons.  The vertical axis measures 
the number of pairs produced in each spatial bin, normalized per primary electron.  
Both curves are for $P = 0.1$ s, $B/B_{\rm cr} = 0.1$,
$T_6 = 0.5$ and colatitude $\xi = 0.7$, in units of PC half-angle.  
    \label{fig:PairSource} }    

\figureout{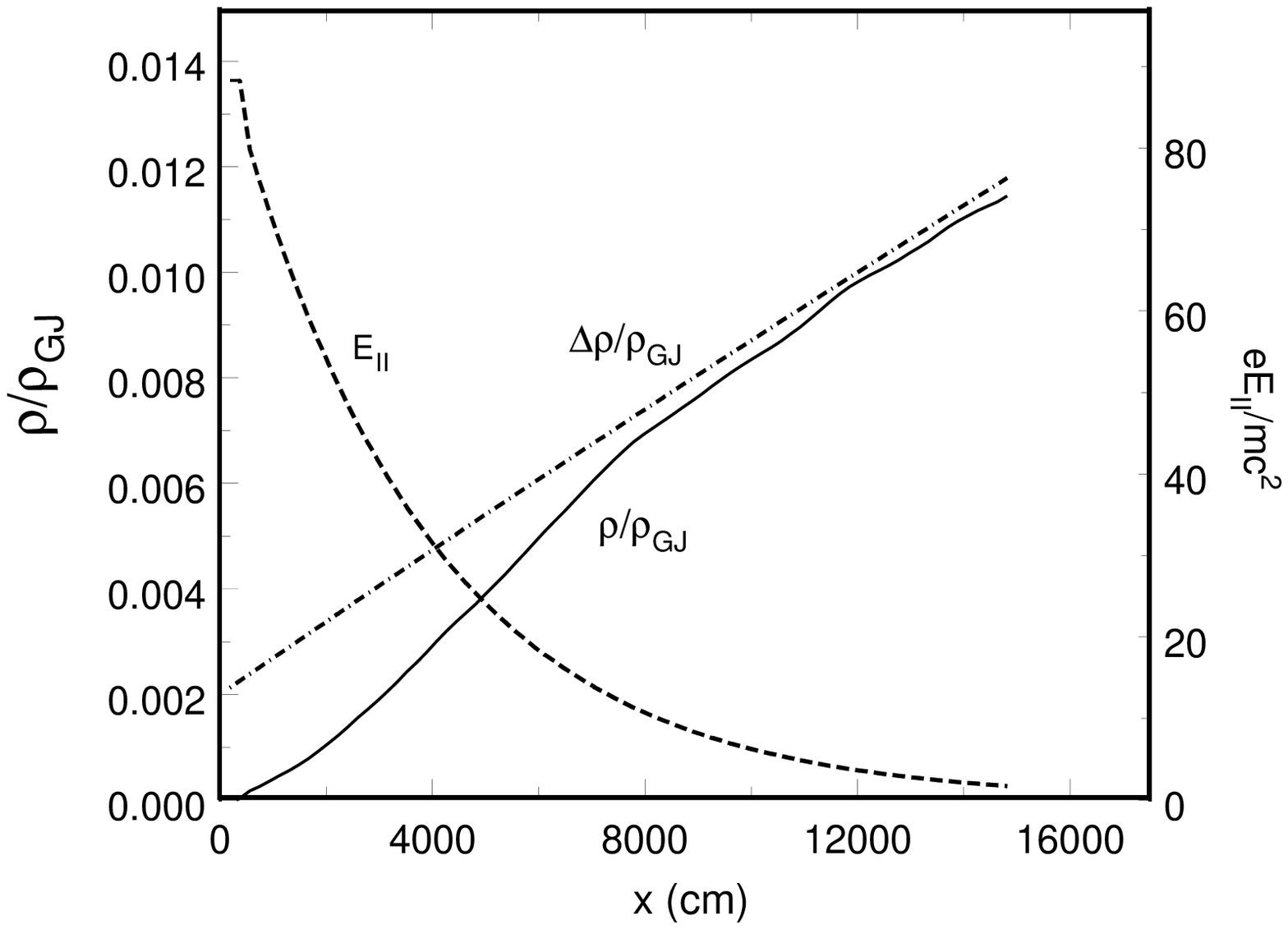}{0}{
Self-consistent solution for the charge density, $\rho(x)$, due to trapped positrons which 
asymptotically approaches the charge deficit, $\Delta \rho$, needed to screen the 
electric field, $E_{\parallel }$, above the PFF, modeled as a declining
exponential with screening scale height, $\Delta_s R$.  $x$ is the distance above the
pair front. The pulsar parameters are the same as those of Figure 2 except that 
$T_6 = 1$ here. 
    \label{fig:rho} }    

\figureout{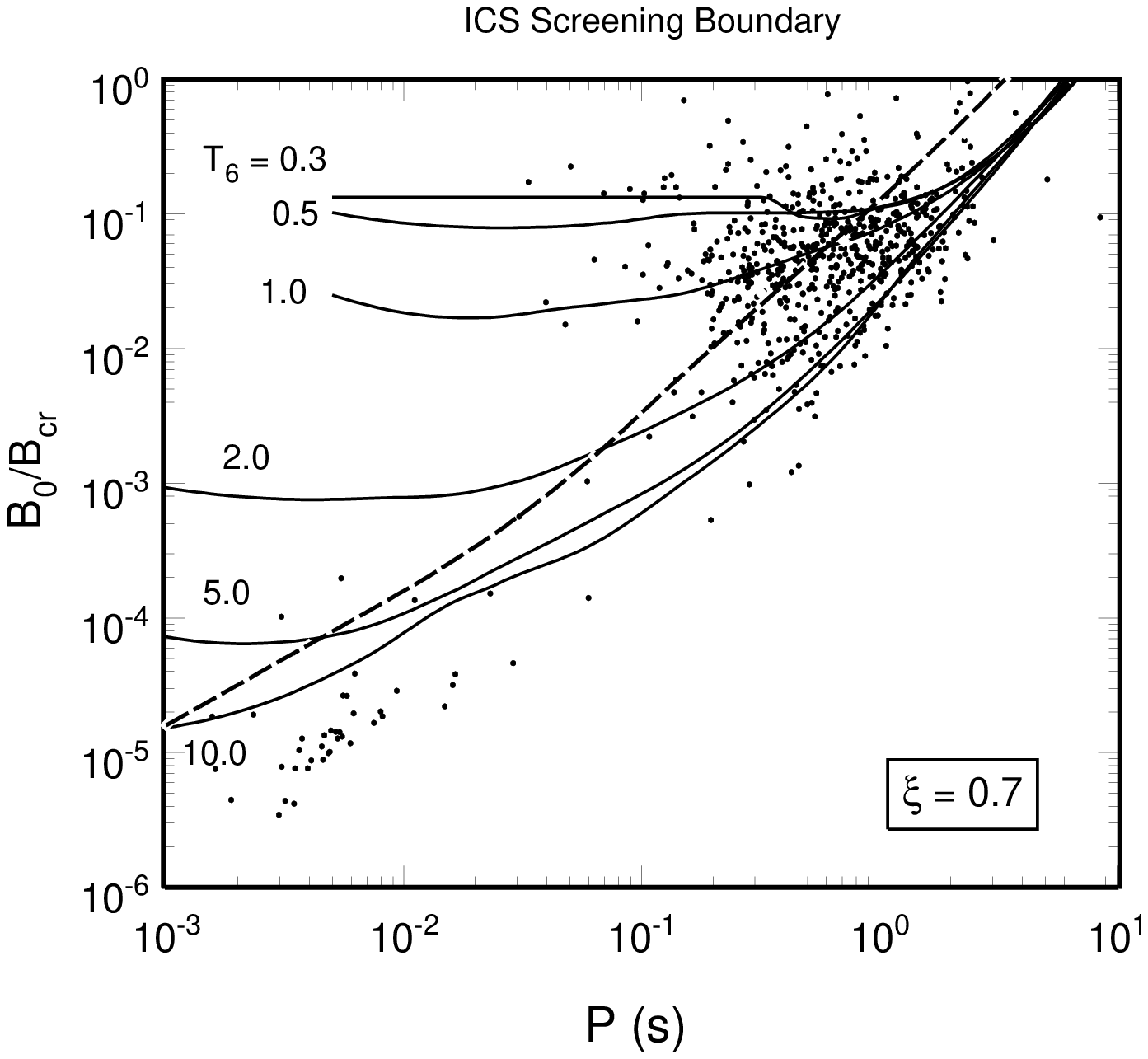}{0}{
Boundaries in surface magnetic field $B_0$ (in units of critical field) 
versus pulsar period, $P$, above which locally complete screening of $E_{\parallel}$
occurs above the ICS pair front.  Curves are labeled with values of the PC 
temperature $T_6$, in units of $10^6$ K.  The dashed curve is the CR pair
boundary (as shown in Figure 1).
    \label{fig:Screen} }    

\figureout{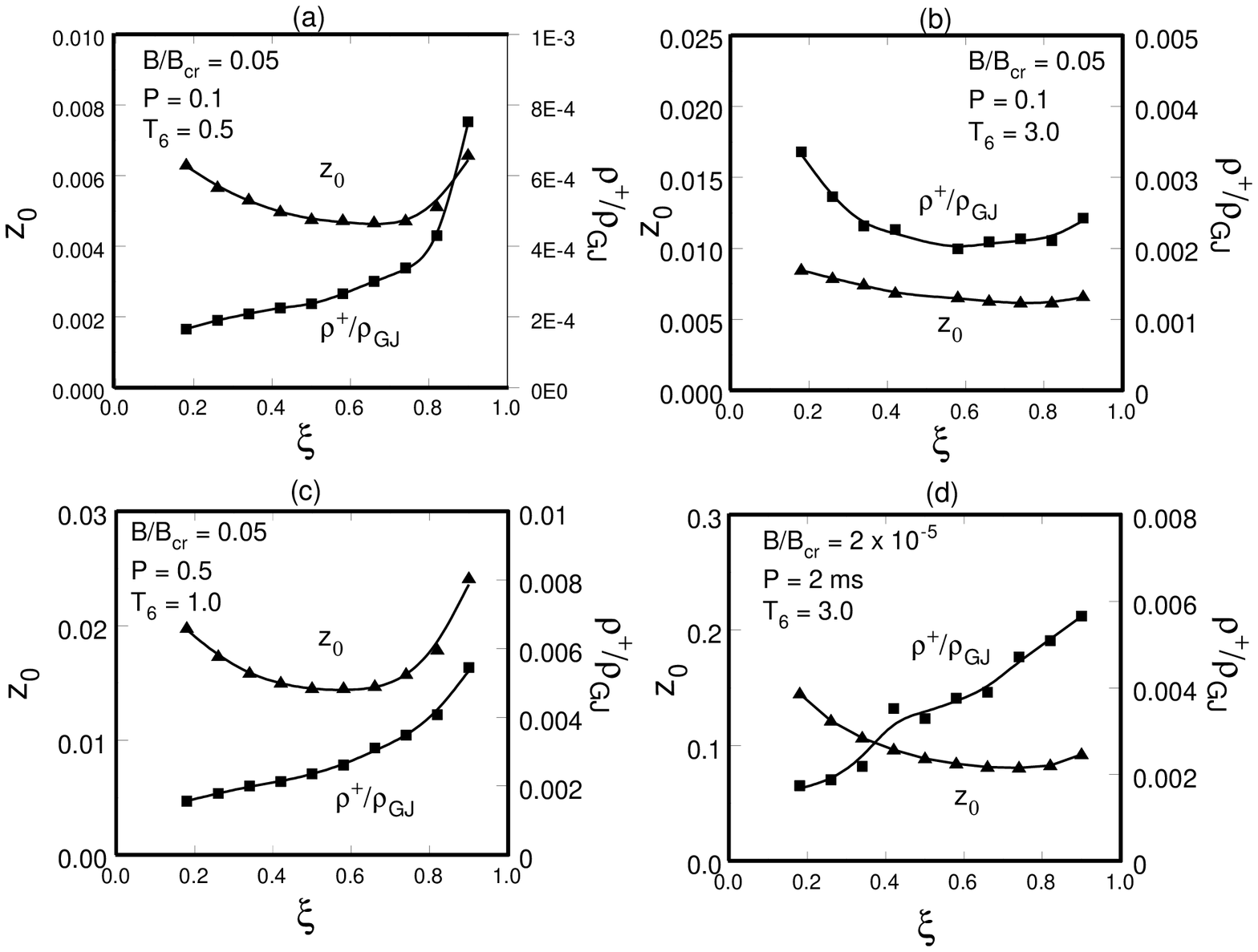}{0}{
Solutions for the returning (trapped) positron density, $\rho^+/\rho _{\rm GJ}$, 
normalized to the Goldreich Julian density and pair formation front altitude, $z_0$,
in units of NS radius, as a function of magnetic colatitude, $\xi$, which has 
been normalized to the PC half-angle.
    \label{fig:rho+_xi} }    

\figureout{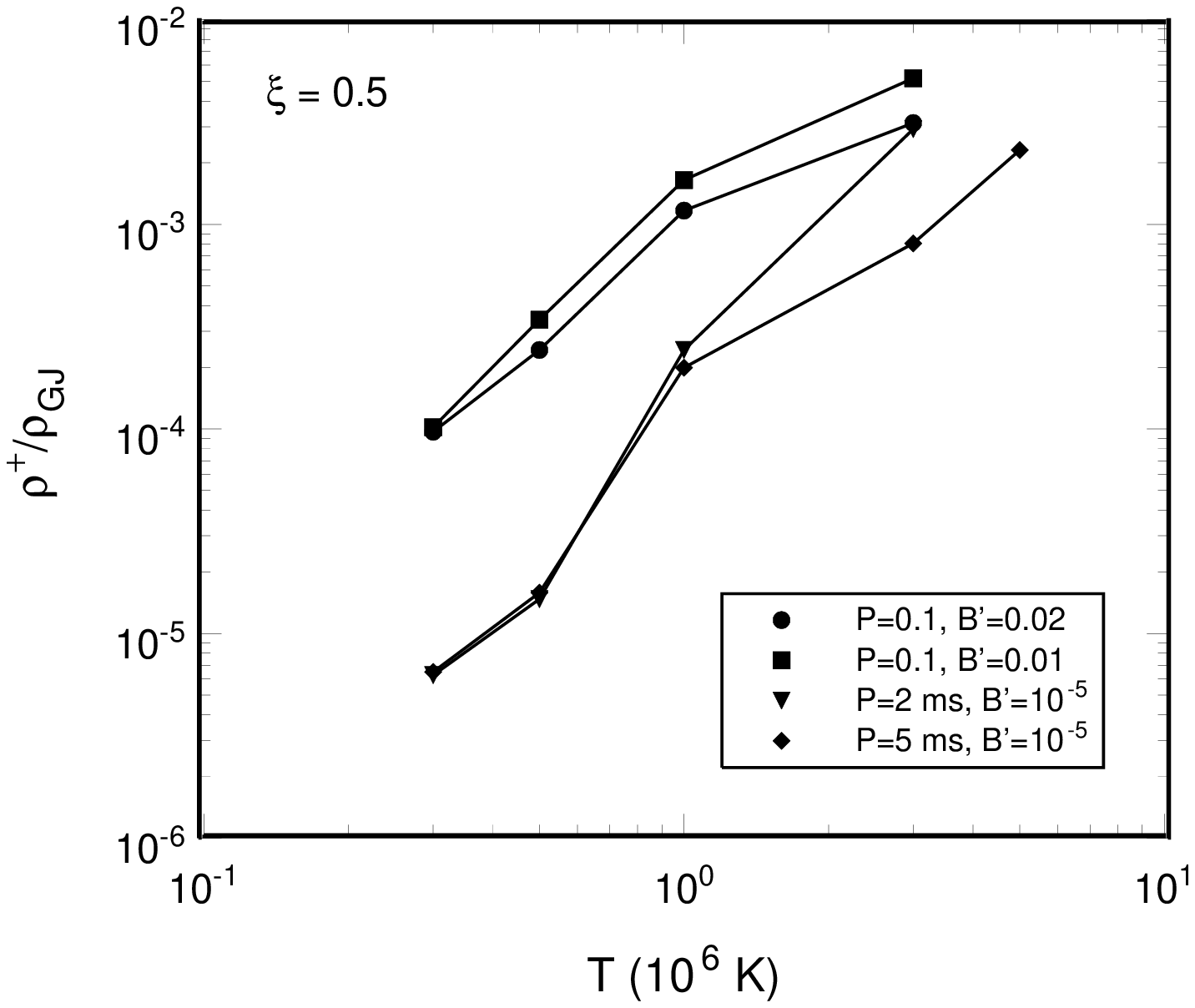}{0}{
Solutions for the returning (trapped) positron density, $\rho^+/\rho_{\rm GJ}$, 
normalized to the Goldreich Julian density as a function of 
PC temperature $T_6$, in units of $10^6$ K, for different values of 
surface magnetic field strength in units of the critical field, $B/B_{\rm cr}$.
    \label{fig:rho+_T} }    

\figureout{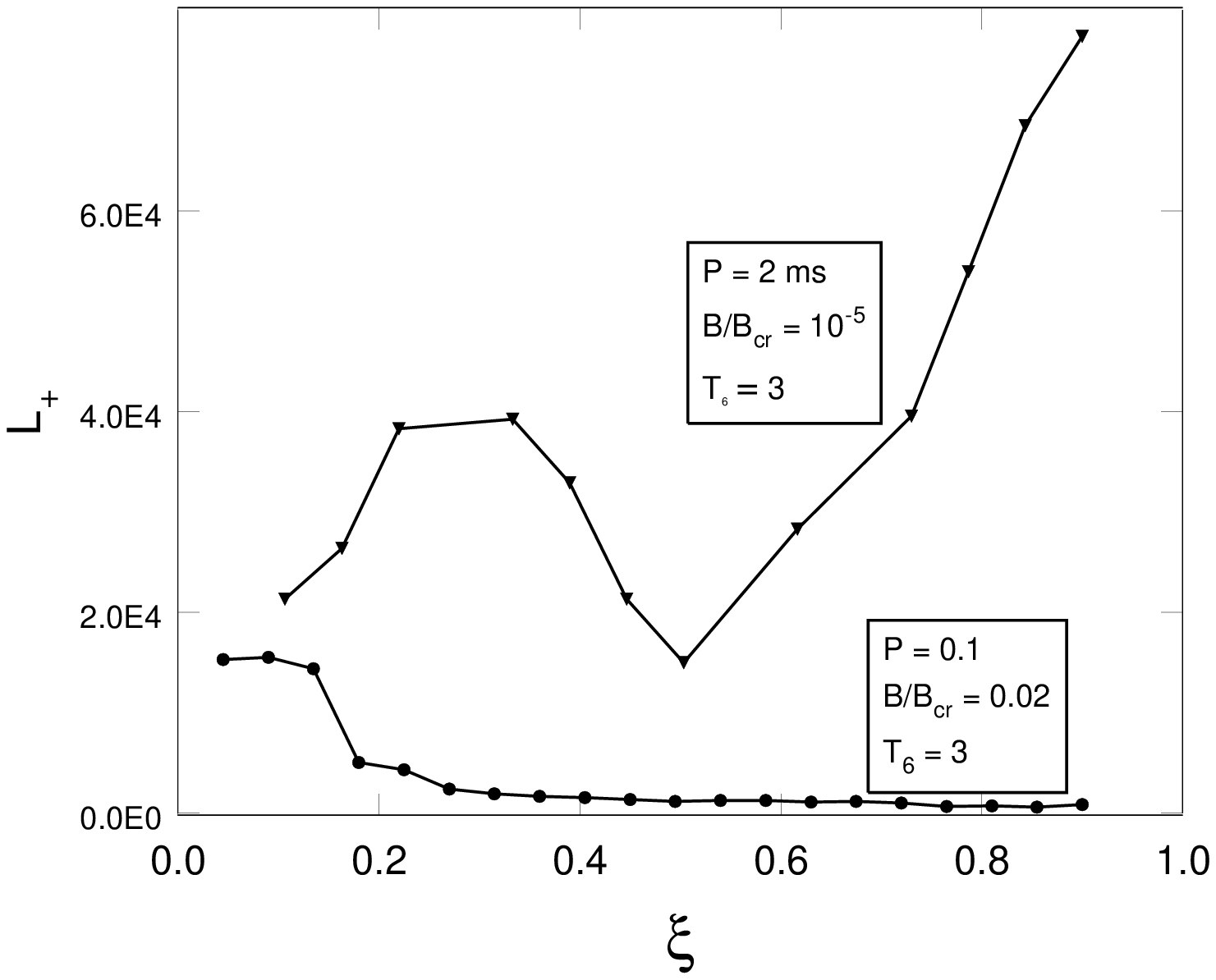}{0}{
Examples of the variation of PC heating luminosity, $L_+$, as a function of  
magnetic colatitude, $\xi$, in units of the PC half-angle.
    \label{fig:L+_xi} }    
 
\figureout{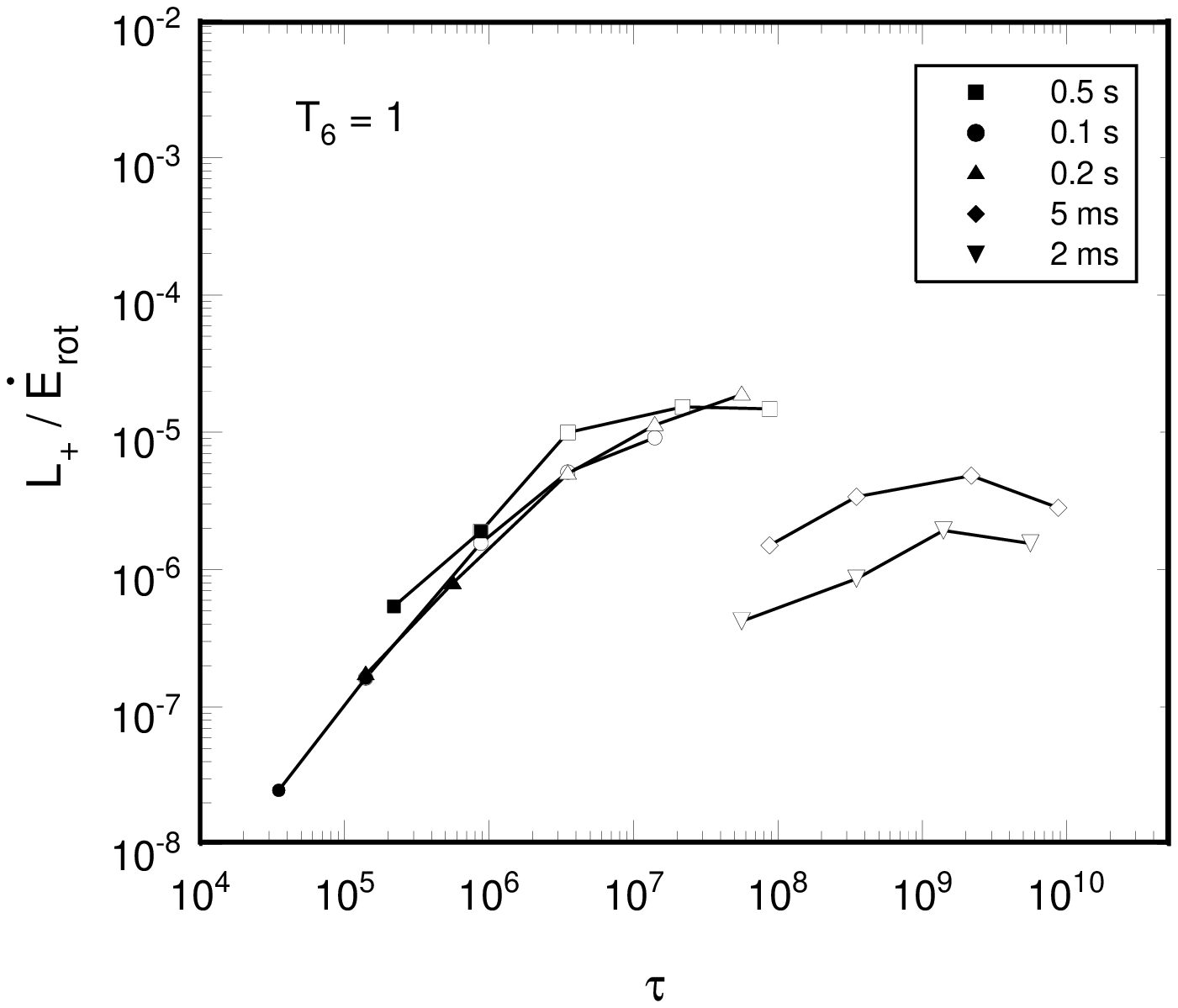}{0}{
PC heating luminosity from ICS pair fronts, $L_+$, 
normalized to the spin-down energy loss rate,
$\dot E_{\rm rot}$, as a function of the characteristic spin-down age, $\tau
= P/2\dot P$, for different pulsar periods, as labeled, and 
PC temperature a) $T_6 = 1.0$ b) $T_6 = 3.0$.  Closed symbols designate 
locally complete screening and open symbols indicate that
no screening occurs above the ICS pair front. 
    \label{fig:L+E_t} }    

\psfig{figure=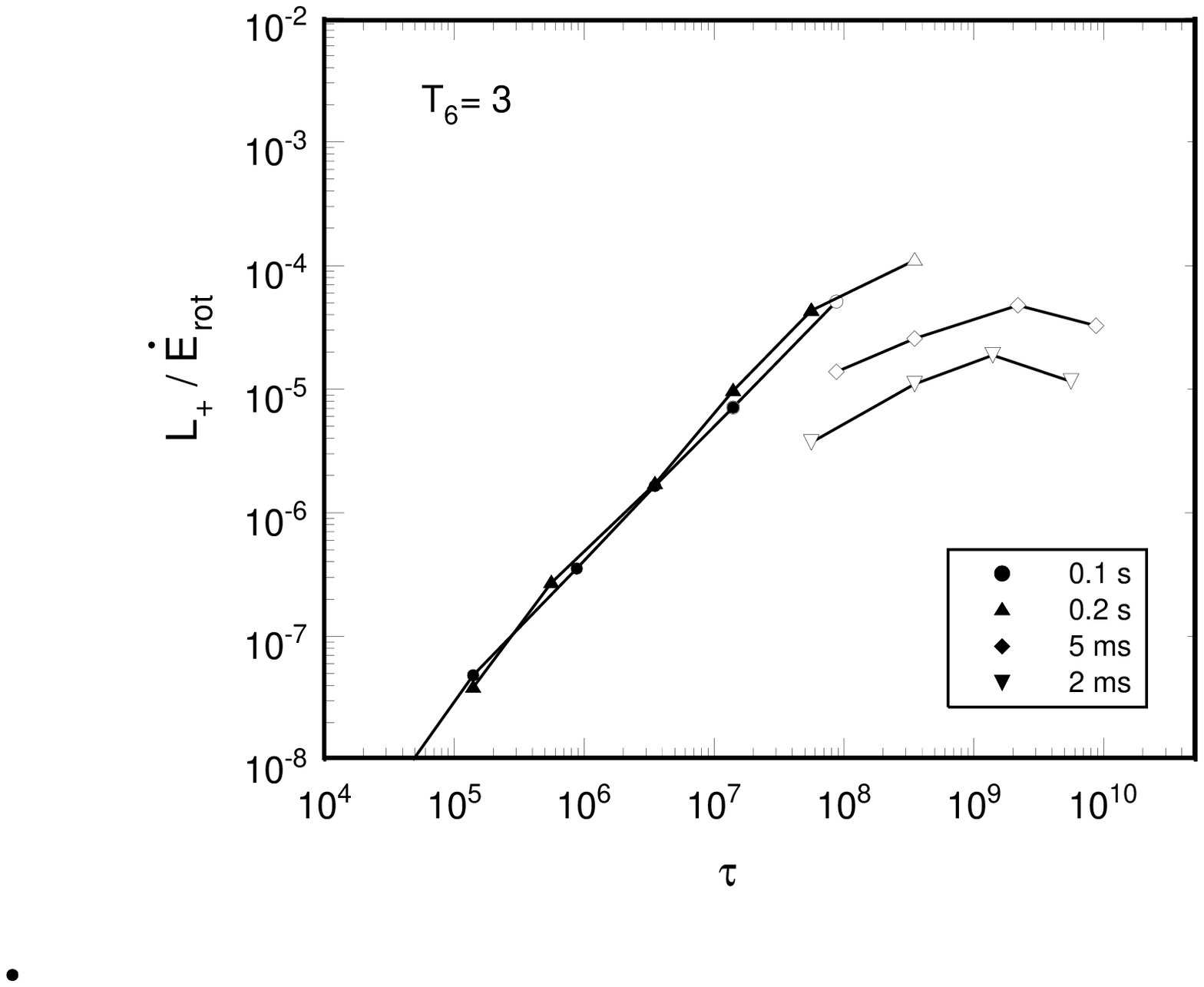,width=7in,angle=0}

\figureout{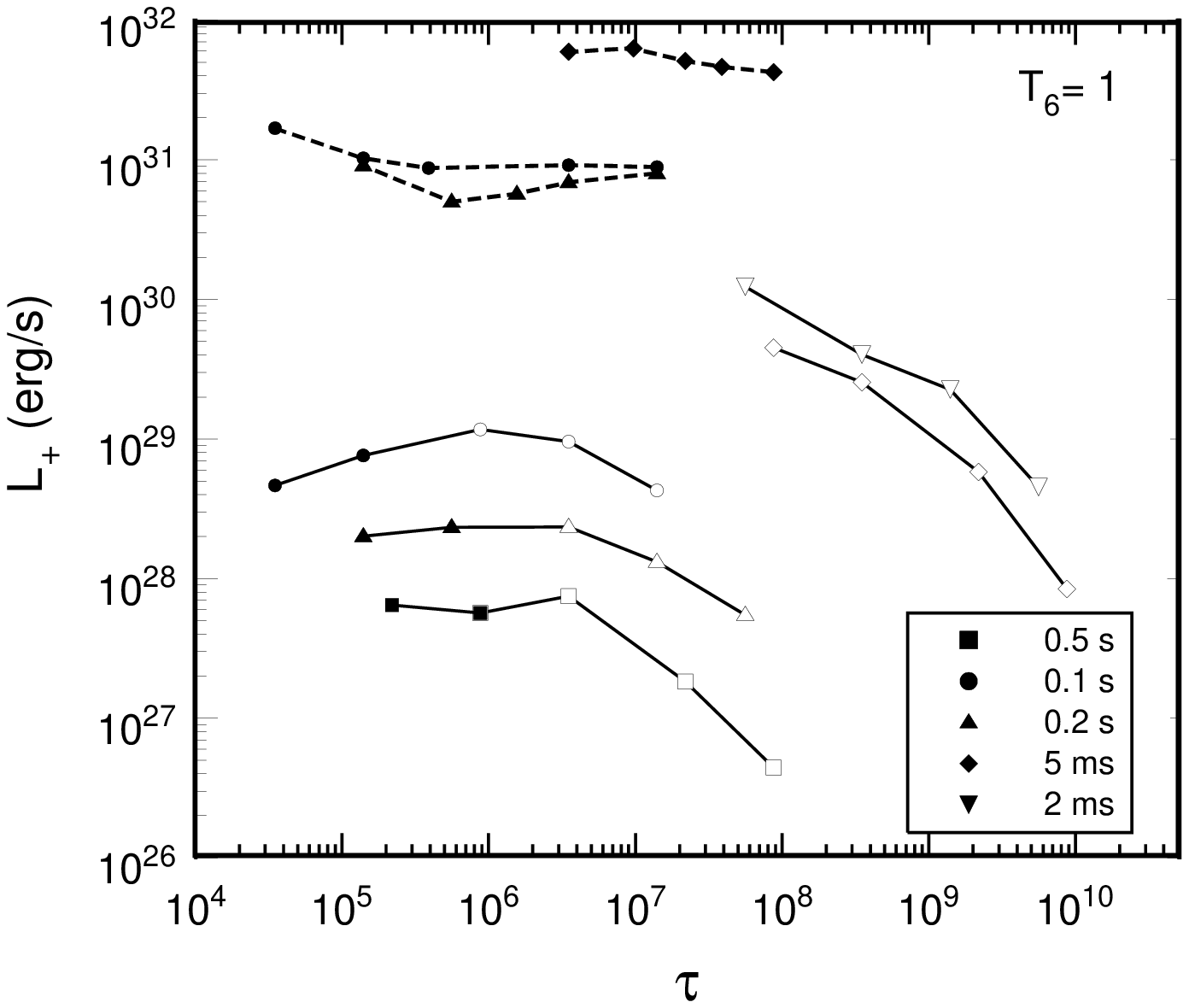}{0}{
PC heating luminosity, $L_+$, as a function of the characteristic 
spin-down age, $\tau = P/2\dot P$, for different pulsar periods, as labeled, 
and PC temperature a) $T_6 = 1.0$ b) $T_6 = 3.0$. 
Solid curves show heating luminosities from ICS pair fronts and dashed curves
show heating luminosities from CR pair fronts (from Paper I). 
Closed symbols designate locally complete screening and open symbols indicate that
no screening occurs above the ICS pair front. 
    \label{fig:L+_t} }    

\psfig{figure=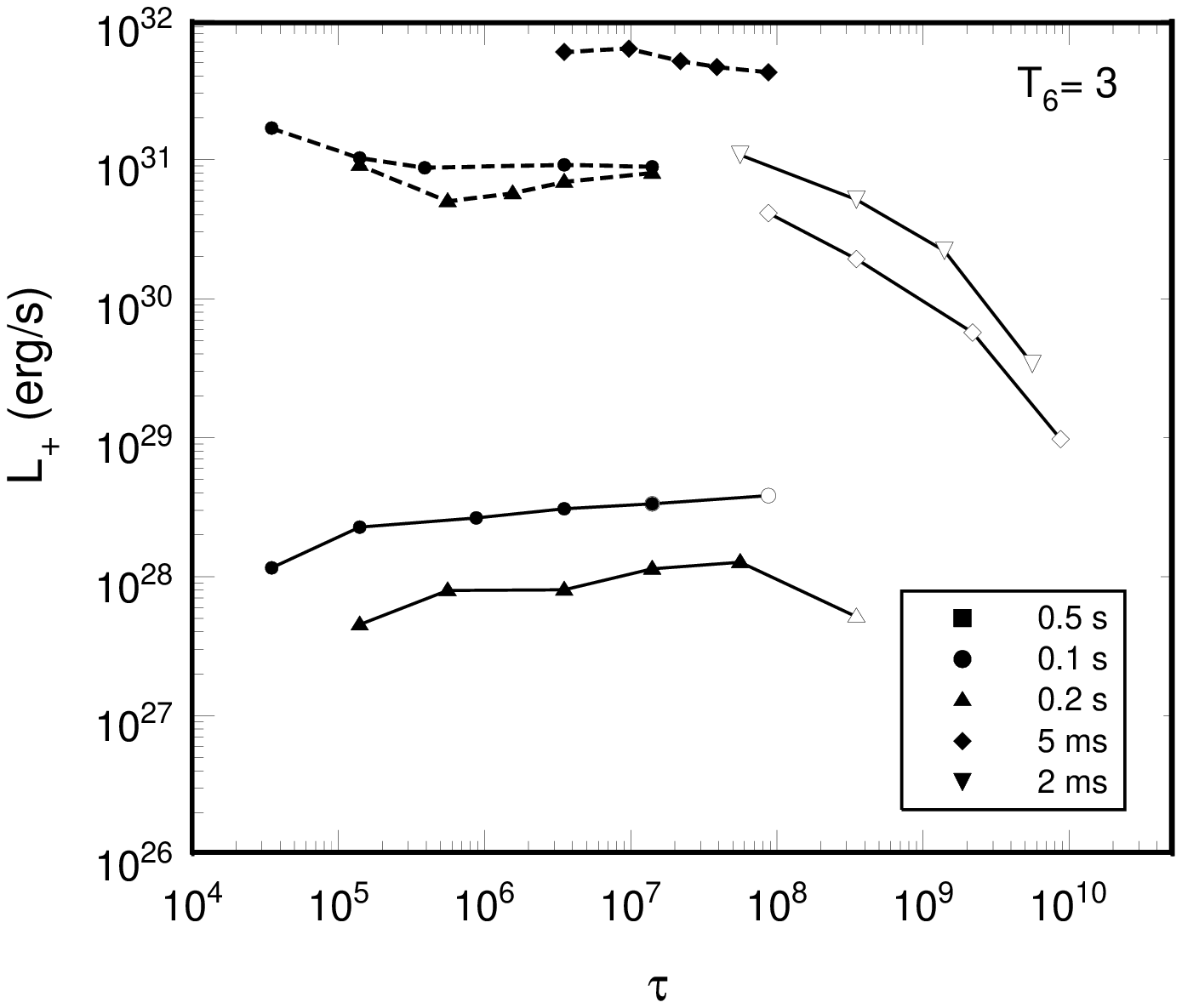,width=7in,angle=0}

\figureout{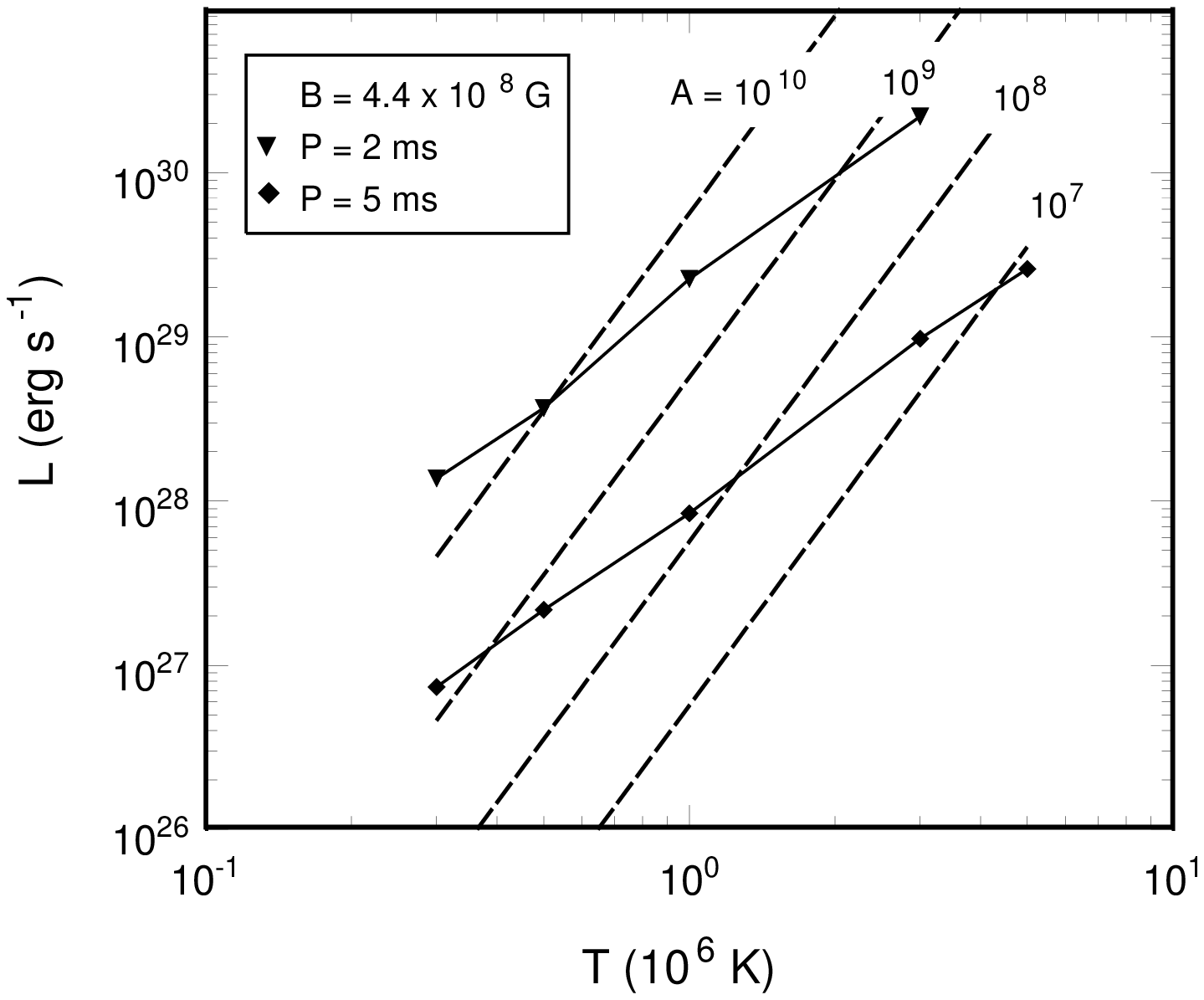}{0}{
PC heating luminosity $L_+$ (solid curves) for different pulsar periods, as
labeled, and blackbody luminosity $L_{BB} = A\sigma _{\rm SB}T^4$ for different areas, $A$
as labeled, as a function of PC temperature, $T$.
    \label{fig:L+_T} }    

\end{document}